# Mancha3D code: Multi-purpose Advanced Non-ideal MHD Code for High resolution simulations in Astrophysics

**M. Modestov[1,2] · E. Khomenko[1,2] · N. Vitas[1,2] · A. de Vicente[1,2] · A. Navarro[1,2] · P. A. González-Morales[1,2] · M. Collados[1,2] · T. Felipe[1,2] · D. Martínez-Gómez[1,2] · P. Hunana[1,2] · M. Luna[3,4] · M. Koll Pistarini[1,2] · B. Popescu Braileanu[5] · A. Perdomo García[1,2] · V. Liakh[5] · I. Santamaria[1,2] · M. M. Gomez Miguez[1,2]**

**Abstract** The Mancha3D code is a versatile tool for numerical simulations of magnetohydrodynamic processes in solar/stellar atmospheres. The code includes non-ideal physics derived from plasma partial ionization, a realistic equation of state and radiative transfer, which allows performing high quality realistic simulations of magneto-convection, as well as idealized simulations of particular processes, such as wave propagation, instabilities or energetic events. The paper summarizes the equations and methods used in the Mancha3D code. It also describes its numerical stability and parallel performance and efficiency. The code is based on a finite difference discretization and memory-saving Runge-Kutta (RK) scheme. It handles non-ideal effects through super-time stepping and Hall diffusion schemes, and takes into account thermal conduction by solving an additional hyperbolic equation for the heat flux. The code is easily configurable to perform different kinds of simulations. Several examples of the code usage are given. It is demonstrated that splitting variables into equilibrium and perturbation parts is essential for simulations of wave propagation in a static background. A perfectly matched layer (PML) boundary condition built into the code greatly facilitates a non-reflective open boundary implementation. Spatial filtering is an important numerical remedy to eliminate grid-size perturbations enhancing the code stability. Parallel

✉ M. Modestov
modestov@iac.es

1 Instituto de Astrofísica de Canarias, 38205 La Laguna, Tenerife, Spain

2 Dpto. de Astrofísica, Universidad de La Laguna, 38205 La Laguna, Tenerife, Spain

3 Departament de Física, Universitat de les Illes Balears, E-07122, Palma de Mallorca, Spain

4 Institute of Applied Computing & Community Code (IAC$^3$), UIB, Spain

5 Centre for Mathematical Plasma Astrophysics, KU Leuven, 3001, Leuven, Belgium

performance analysis reveals that the code is strongly memory bound, which is a natural consequence of the numerical techniques used, such as split variables and PML boundary conditions. Both strong and weak scalings show adequate performance up till several thousands of CPUs.

## 1. Introduction

Stellar interiors and atmospheres are intrinsically multi-scale environments. Subject to gravitational force, the plasma parameters are strongly stratified with stellar radial distance. Stellar plasmas are dynamic on a wide variety of scales, ranging from scales of short-lived transients and high-frequency phenomena associated to the energy dissipation (high-frequency waves, shock wave fronts, current layers, reconnection sites, etc), to long scales associated to the time of evolution of supergranulation, active regions, or stellar rotation. Theoretical-analytical studies of such environments are only practical in some limited cases. Already for decades, numerical simulations have been a useful tool for modeling complex plasma interactions, making it possible to obtain a global and precise picture of the dynamics and energy propagation through stellar interiors and atmospheres.

Nowadays a large number of computer codes are used for various problems of magnetized plasma dynamics. Some of these codes are generic-purpose codes, as, for example, PENCIL (Collaboration *et al.*, 2021), MPI-AMRVAC (Xia *et al.*, 2018; Keppens *et al.*, 2023), Athena (Stone *et al.*, 2008) or FLASH (Fryxell *et al.*, 2000). Other codes are specifically designed to model particular problems. For example, in the area of astrophysics, several groups have advanced codes to perform realistic simulations of magneto-convection in solar/stellar atmospheres, such as CO5BOLD (Freytag, 2013), MURaM (Vögler *et al.*, 2005), Bifrost (Gudiksen *et al.*, 2011), SolarBox (Wray *et al.*, 2015), RAMENS (Iijima and Yokoyama, 2015), or Mancha3D (Khomenko *et al.*, 2018, discussed in this paper). Many of the realistic magneto-convection models extend from deep sub-photospheric layers till high up in the corona (Carlsson *et al.*, 2016; Rempel, 2017; Iijima and Yokoyama, 2017). In this regard, Bifrost provided one of the first self-consistent models of solar coronal heating by Ohmic dissipation, where the role of Ohmic dissipation was played by hyper-diffusion terms (Gudiksen and Nordlund, 2005).

There is a group of codes that have been particularly configured to perform realistic large-scale stellar modeling of the dynamics of the convection zone, dynamos and differential rotation, see very recent applications by PENCIL (Käpylä, 2023), R2D2 (Hotta, Kusano, and Shimada, 2022), ASH (Brun *et al.*, 2022), or RAMSES (Canivete Cuissa and Teyssier, 2022). This kind of simulations frequently uses their own set of assumptions, such as reduced sound speed technique, inelastic approximation, etc.

In stellar atmospheres, the interaction of plasma and radiation plays a crucial role, with radiation being the main cooling mechanism. Realistic codes include advanced radiative transfer calculations, integrating the solution of the radiative transfer equation under several assumptions that allow to speed up the calculations. Among the various codes, only a few take into account departures from local thermodynamic equilibrium (LTE) or time-dependent ionization (Gudiksen *et al.*, 2011; Przybylski *et al.*, 2022). Some recently implemented features consist in the inclusion of non-ideal terms derived from the presence of a large amount of neutrals in the solar atmosphere (ambipolar diffusion, modified Hall term, Biermann battery term) (Cheung and Cameron, 2012; Martínez-Sykora, De Pontieu, and Hansteen, 2012; Khomenko and Collados, 2012). In this regard, MANCHA3D was one of the first codes to take these effects into account for realistic 3D solar magneto-convection simulations (Khomenko *et al.*, 2018), suggesting they could play a relevant role for chromospheric heating.

The codes mentioned above use a single-fluid quasi-magnetohydrodynamic (MHD) formalism. Several newer codes have implemented a multi-fluid formalism for solar plasmas, where the different plasma components (such as neutral and charged components) are evolved separately as independent fluids interacting by collisions (Hillier, Takasao, and Nakamura, 2016; Lukin *et al.*, 2016; Martínez-Gómez, Soler, and Terradas, 2017; Lani *et al.*, 2017; Popescu Braileanu *et al.*, 2019; Wójcik, Murawski, and Musielak, 2019; Martínez-Sykora *et al.*, 2020; Popescu Braileanu and Keppens, 2022). This different class of codes still lack realistic physics, and will not be discussed here. The single-fluid formalism is conceptually easier since the numerical methods for the solution of the MHD system of equations have been well developed. Nevertheless, ambipolar and Hall terms require a special treatment to assure the numerical stability and to speed up the calculations (Arber *et al.*, 2001; Tóth, Ma, and Gombosi, 2008; González-Morales *et al.*, 2018; Nóbrega-Siverio *et al.*, 2020; Popescu Braileanu and Keppens, 2021).

While realistic simulations have proven to be an extremely useful tool, their complexity is frequently similar to actual observations, making it difficult to disentangle between different physical processes to explain particular phenomena. A different class of simulations is frequently used instead, classified as idealized ones. In idealized simulations, one is only interested in performing a controlled experiment with a limited number of the physical ingredients. Simulations of waves, propagating through a static background (as, e.g., Hasan *et al.*, 2003; Vigeesh, Hasan, and Steiner, 2009; Bogdan *et al.*, 2003; Fedun, Shelyag, and Erdélyi, 2011; Santamaria, Khomenko, and Collados, 2015; Liakh, Luna, and Khomenko, 2021, to name a few) and instabilities in stellar atmospheres (e.g., Terradas *et al.*, 2008; Hillier *et al.*, 2012; Antolin, Yokoyama, and Van Doorsselaere, 2014) are examples of this kind of simulations. Idealized simulations have been helpful in answering questions about wave energy transfer and dissipation through solar/stellar atmospheres, chromospheric heating by waves and instabilities (see recent reviews by Hillier, 2018; Srivastava *et al.*, 2021).

As follows from the discussion above, many algorithms and codes for astrophysical plasma simulations are available, with a variety of numerical implementations. Our primary goal in developing MANCHA3D[1] is to produce a versatile code that is easily configurable for simulations with idealized and controlled setups to realistic simulations, such as the one of solar/stellar magneto-convection. Therefore, the code presents a modular structure and a user switches on/off a given functionality with a number of pre-processor commands. The ideal and non-ideal modules of the code share the same core integration scheme. A user does not need to modify the core of the code to switch between the setups, but can do it through external initial and boundary condition modules. MANCHA3D is not the first or only code that uses its particular numerical algorithms, but according to our experience, precise implementation details are important and produce differences between the codes, in a more or less substantial manner. Therefore, MANCHA3D has been designed to be an easy to use, configure, develop, and maintain code, which can be expanded to add more physics. The code solves the time-dependent equations of the non-ideal MHD on a 3D Cartesian grid, and it can solve either full or linearized MHD equations. The code is written in modern FORTRAN language, fully parallelized with MPI, using spatial domain decomposition; input/output files are handles via parallel HDF5.

[1]The name MANCHA3D goes back to the earliest version of the code that was designed to simulate propagation of waves in sunspots ("mancha solar" means sunspot in Spanish). Written in all capitals, it is an acronym for Multi-fluid (-purpose, -physics, -dimensional) Advanced Non-ideal MHD Code for High resolution simulations of the solar Atmosphere.

The original version of the code solver and algorithm has been described and validated with multiple tests in Felipe, Khomenko, and Collados (2010). Compared to that work, many crucial changes have been done to the code, which justifies the need for the current publication. Several optimizations of the modular structure, efficiency and parallelization have been performed. Realistic modules have been added for calculations using a tabulated equation of state (EOS), accounting for the chemical composition of a stellar atmosphere (Vitas and Khomenko, 2015), see also Perdomo García *et al.* (2023), as well as a module for solving the radiative transfer equation (RTE) in LTE approximation using opacity binning (partially described in Khomenko *et al.*, 2018). Non-ideal processes such as ambipolar diffusion (caused by the presence of neutrals) and the Hall and Biermann battery effects have been included using time efficient numerical methods such as super-time-stepping (STS) or Hall diffusion scheme (HDS) (González-Morales *et al.*, 2018). Together with the radiative transfer module, it allows performing realistic simulations of solar/stellar atmospheres to be directly compared with observations. The recent implementation of a thermal conduction module enables extending our modeling to the corona (Navarro *et al.*, 2022). Furthermore, unlike many specialized astrophysical codes, MANCHA3D is freely available via public Gitlab repository, https://gitlab.com/Mancha3D.

Starting from the studies of magneto-acoustic and Alfvén waves in sunspots (Khomenko and Collados, 2006, 2009; Felipe, Khomenko, and Collados, 2010, 2011; Krishna Prasad, Jess, and Khomenko, 2015; Zhao *et al.*, 2016), and other magnetic structures (Khomenko, Collados, and Felipe, 2008; Santamaria, Khomenko, and Collados, 2015; Santamaria and Van Doorsselaere, 2018; Riedl, Van Doorsselaere, and Santamaria, 2019; Sieyra *et al.*, 2022), the MANCHA3D code has successfully been applied for different setups including thorough investigation of non-ideal MHD effects due to neutrals (Khomenko and Collados, 2012; Khomenko *et al.*, 2014b; Shelyag *et al.*, 2016; MacBride *et al.*, 2022), realistic magneto-convection of the solar atmosphere (Khomenko *et al.*, 2017, 2018; González-Morales *et al.*, 2020), large-amplitude oscillations of solar prominences (Luna *et al.*, 2016; Liakh, Luna, and Khomenko, 2020, 2021, 2023; Luna and Moreno-Insertis, 2021), or solar seismology (Felipe *et al.*, 2016; Felipe, Braun, and Birch, 2017; Felipe *et al.*, 2020). All these works have proved the code capability and its competence to solve complex physical problems, however many technical aspects of the MANCHA3D code have not been reported. The aim of this paper is to summarize the capabilities of the code, describe its properties, together with accurate analyses of its numerical stability and efficiency. The paper is organized as follows. Section 2 presents the equations used in the code and Sect. 3 the numerical methods and stabilization techniques. Sections 2.6 and 3.7 emphasize the importance of the split variables and the PML boundary conditions. Parallel efficiency is thoroughly studied and described in Section 4. The paper is completed with the conclusions and future plans in Section 5.

## 2. Equations

The Mancha3D code solves the following non-ideal MHD equations, written in conservative form[2]:

$$\frac{\partial \rho}{\partial t} + \nabla \cdot (\rho \mathbf{v}) = \left(\frac{\partial \rho}{\partial t}\right)_{\mathrm{diff}}, \tag{1}$$

$$\frac{\partial \rho \mathbf{v}}{\partial t} + \nabla \cdot \left[\rho \mathbf{v}\mathbf{v} + \left(p + \frac{\mathbf{B}^2}{2\mu_0}\right)\mathbf{I} - \frac{\mathbf{B}\mathbf{B}}{\mu_0}\right] = \rho \mathbf{g} + \mathbf{S}(t) + \left(\frac{\partial \rho \mathbf{v}}{\partial t}\right)_{\mathrm{diff}}, \tag{2}$$

$$\begin{aligned} \frac{\partial e_{\mathrm{tot}}}{\partial t} + \nabla \cdot \Bigg[\mathbf{v}\left(e_{\mathrm{tot}} + p + \frac{\mathbf{B}^2}{2\mu_0}\right) - \frac{\mathbf{B}(\mathbf{v}\cdot\mathbf{B})}{\mu_0} + \frac{(\eta_{\mathrm{A}} + \eta)\mathbf{J}_\perp \times \mathbf{B}}{\mu_0} \\ - \frac{\nabla p_{\mathrm{e}} \times \mathbf{B}}{e n_{\mathrm{e}} \mu_0} + \mathbf{q}\Bigg] = (\rho \mathbf{g} + \mathbf{S}(t)) \cdot \mathbf{v} + Q_{\mathrm{R}} + \left(\frac{\partial e_{\mathrm{tot}}}{\partial t}\right)_{\mathrm{diff}}, \end{aligned} \tag{3}$$

$$\frac{\partial \mathbf{B}}{\partial t} = \nabla \times \left[\mathbf{v} \times \mathbf{B} - \eta \mathbf{J} - \eta_{\mathrm{A}} \mathbf{J}_\perp + \frac{\nabla p_{\mathrm{e}}}{e n_{\mathrm{e}}} - \eta_{\mathrm{H}} \frac{(\mathbf{J} \times \mathbf{B})}{|\mathbf{B}|}\right] + \left(\frac{\partial \mathbf{B}}{\partial t}\right)_{\mathrm{diff}}, \tag{4}$$

where $\rho$ is the density, $\mathbf{v}$ is the velocity, $p$ is the gas pressure, $p_{\mathrm{e}}$ is the electron pressure (see Sect. 2.5.3), $n_{\mathrm{e}}$ is the electron number density, $\mathbf{B}$ is the magnetic field, $\mathbf{I}$ is the identity tensor, $\mathbf{g}$ is the gravitational acceleration, $\mathbf{q}$ is the heat flux vector, and the non-ideal terms are described later in this section. The dot '$\cdot$' represents the scalar product, while the notation '$\mathbf{BB}$' (or '$\mathbf{vv}$') stands for the tensor product. The term $\mathbf{S}(t)$ in the momentum and energy equations represents a time-dependent external force. The term $Q_{\mathrm{R}}$ stands for the radiative energy exchange. The total electric current, and the current perpendicular to the magnetic field are defined as

$$\mathbf{J} = \frac{\nabla \times \mathbf{B}}{\mu_0}, \tag{5}$$

$$\mathbf{J}_\perp = -\frac{(\mathbf{J} \times \mathbf{B}) \times \mathbf{B}}{\mathbf{B}^2}. \tag{6}$$

Artificial diffusion terms, marked as $()_{\mathrm{diff}}$, have been added to the conservation equations for the stability of the simulations. The diffusion terms in the momentum, energy and induction equations have their physical counterparts, but the one in the continuity equation does not. These terms will be described in more detail below.

[2]Everywhere in the paper we use the international SI units.

The energy conservation equation is written in terms of the total energy per unit volume,

$$e_{\rm tot} = e_{\rm int} + \frac{1}{2}\rho \mathbf{v}^2 + \frac{\mathbf{B}^2}{2\mu_0}, \tag{7}$$

where $e_{\rm int}$ is the internal energy per unit volume that is defined by the equation of state (see Sect. 2.5). Alternatively to the total energy equation, Mancha3D can solve the equation of conservation of the internal energy,

$$\frac{\partial e_{\rm int}}{\partial t} + \nabla \cdot (\mathbf{v} e_{\rm int} + \mathbf{q}) + p \nabla \cdot \mathbf{v} = \\ \eta J^2 + \eta_{\rm A} {J_\perp}^2 - \mathbf{J} \cdot \frac{\nabla p_{\rm e}}{e n_{\rm e}} + Q_{\rm R} + \left( \frac{\partial e_{\rm int}}{\partial t} \right)_{\rm diff} . \tag{8}$$

When the magnetic pressure is much larger than the gas pressure, recovering thermal energy ($e_{\rm int}$) or pressure from the total energy ($e_{\rm tot}$) can lead to numerical errors. Hence, in the runs where a large portion of the simulation domain contains plasma with $\beta \ll 1$, Mancha3D can be set to use the equation of conservation of the internal energy, Eq. 8.

In order to handle small densities in a numerically stable way, the continuity equation can be solved in terms of logarithmic density, $\varphi \equiv \ln \rho$,

$$\frac{\partial \varphi}{\partial t} + \nabla \cdot \mathbf{v} + \mathbf{v} \cdot \nabla \varphi = \nabla \cdot (\nu \nabla \varphi) + \nu (\nabla \varphi)^2, \tag{9}$$

where $\nu$ is the artificial diffusivity coefficient, defined in Eq. 82 for the linear density, see Sect. 3.3.

## 2.1. Treatment of the heat conduction

The effects of thermal conduction can be represented by adding the divergence of a heat flux vector $\mathbf{q}$ either to the right hand side of Eq. 3,

$$\frac{\partial e_{\rm tot}}{\partial t} = [\dots] - \nabla \cdot \mathbf{q}\,, \tag{10}$$

or to the right hand side of Eq. 8 if the internal energy is evolved instead of the total energy. The classical heat flux description (with expansions around a Maxwellian distribution function) was addressed for example by Braginskii (1965); Spitzer (1956); Schunk (1977); Balescu (1988); Zhdanov (2002); Hunana *et al.* (2022) and references therein. For general HD and MHD problems, the conductivities can be set to constant values for isotropic and anisotropic cases. With respect to the magnetic field the heat flux $\mathbf{q} = -\kappa \nabla T$ can be decomposed as follows

$$\mathbf{q} = -\kappa_{\|} \nabla_{\|} T - \kappa_{\perp} \nabla_{\perp} T + \kappa_{\times} \hat{\mathbf{b}} \times \nabla_{\perp} T\,, \tag{11}$$

where $\nabla_{\|} = \hat{\mathbf{b}}(\hat{\mathbf{b}} \cdot \nabla)$ gives the parallel projection to the magnetic field, $\nabla_{\perp} = \nabla - \nabla_{\|}$ gives the projection in the perpendicular direction, and the last term is the projection in the transverse direction.

The model derived by Braginskii (1965) takes into account the dependency on the ratio of the collisional to cyclotron frequencies of the plasma. The model transitions smoothly between field-aligned conductivity and isotropic conductivity for regions with a low or zero magnetic field or collisionally dominated. The joint contribution from ions and electrons is

$$\kappa_{\|} = \kappa_{\|}^{\rm e} + \kappa_{\|}^{\rm i} , \quad (12)$$

$$\kappa_{\perp} = \kappa_{\perp}^{\rm e} + \kappa_{\perp}^{\rm i} , \quad (13)$$

$$\kappa_{\times} = \kappa_{\times}^{\rm e} + \kappa_{\times}^{\rm i} , \quad (14)$$

where the lower and upper indices "e" and "i" refer to electrons and ions. The conductivities are given by

$$\kappa_{\|}^{\rm e} = 3.1616 \frac{k_{\rm B} p_{\rm e}}{\nu_{\rm ei} m_{\rm e}} , \quad (15)$$

$$\kappa_{\perp}^{\rm e} = \frac{k_{\rm B} p_{\rm e}}{\nu_{\rm ei} m_{\rm e}} \frac{4.664 x_{\rm e}^2 + 11.92}{x_{\rm e}^4 + 14.79 x_{\rm e}^2 + 3.77} , \quad (16)$$

$$\kappa_{\times}^{\rm e} = \frac{k_{\rm B} p_{\rm e}}{\nu_{\rm ei} m_{\rm e}} x_{\rm e} \frac{\frac{5}{2} x_{\rm e}^2 + 21.67}{x_{\rm e}^4 + 14.79 x_{\rm e}^2 + 3.77} , \quad (17)$$

$$\kappa_{\|}^{\rm i} = 3.906 \frac{k_{\rm B} p_{\rm i}}{\nu_{\rm ii} m_{\rm i}} , \quad (18)$$

$$\kappa_{\perp}^{\rm i} = \frac{k_{\rm B} p_{\rm i}}{\nu_{\rm ii} m_{\rm i}} \frac{2 x_{\rm i}^2 + 2.645}{x_{\rm i}^4 + 2.70 x_{\rm i}^2 + 0.677} , \quad (19)$$

$$\kappa_{\times}^{\rm i} = \frac{k_{\rm B} p_{\rm i}}{\nu_{\rm ii} m_{\rm i}} x_{\rm i} \frac{\frac{5}{2} x_{\rm i}^2 + 4.65}{x_{\rm i}^4 + 2.70 x_{\rm i}^2 + 0.677} . \quad (20)$$

where $p_{\rm e}$, $p_{\rm i}$, $m_{\rm e}$ and $m_{\rm i}$ are the pressure and mass of each species and $k_{\rm B}$ is the Boltzmann constant. The collisional frequency of ion-ion collisions is $\nu_{\rm ii}$ and of electron-ion collisions is $\nu_{\rm ei}$. The quantities $x_{\rm e}$ and $x_{\rm i}$ represent the ratio between cyclotron frequency ($\Omega$) and collision frequency ($\nu$) for electrons and ions, respectively

$$x_{\rm e} = \frac{\Omega_{\rm e}}{\nu_{\rm ei}}; \qquad x_{\rm i} = \frac{\Omega_{\rm i}}{\nu_{\rm ii}} , \quad (21)$$

and the collision frequencies are specified in Sect. 2.2, see also (Khomenko *et al.*, 2014b). The Braginskii model recovers Spitzer's expression for electron heat flux in the strongly magnetized limit. In the Spitzer's case the constant is 3.203 instead of 3.1616 in Eq. (15).

The thermal conduction module in Mancha3D offers different heat flux models that can be solved with two alternative numerical schemes, it is thoroughly described in Navarro *et al.* (2022). The first option involves the standard explicit integration of the equations including the parabolic term ($\nabla \cdot \mathbf{q}$) in the energy equation. However, in conditions of high plasma temperatures, like the ones in the solar corona, the integration time step due to the thermal conduction becomes considerably more restrictive than the MHD time step and simulations can be computationally expensive. To overcome this difficulty, the second scheme resolves an additional hyperbolic equation for the parallel component of the heat flux, which is the problematic one in the case of astrophysical plasma. This allows the use of larger values of the time step and reduces the cost of the simulations. To do this we rewrite the heat flux

$$\mathbf{q} = -q_{\parallel}\hat{\mathbf{b}} - \kappa_{\perp}\nabla_{\perp}T + \kappa_{\times}\hat{\mathbf{b}} \times \nabla_{\perp}T \,, \tag{22}$$

and introduce a hyperbolic equation for its evolution,

$$\frac{\partial q_{\parallel}}{\partial t} = \frac{1}{\tau}\left(-f_{\mathrm{sat}}\kappa_{\parallel}\left(\hat{\mathbf{b}} \cdot \nabla\right) T - q_{\parallel}\right). \tag{23}$$

The factor $f_{\mathrm{sat}}$ sets the saturation of the conductive heat flux, reducing its value for stability purposes, and $\tau$ is the relaxation time set to $\tau = 4dt$. Following Fisher, Canfield, and McClymont (1985) and Meyer, Balsara, and Aslam (2012) the saturation factor is written as

$$f_{\mathrm{sat}} = \left(1 + \frac{|\kappa_{\parallel}\left(\hat{\mathbf{b}} \cdot \nabla\right) T|}{1.5\rho \mathrm{c}_{\mathrm{S}}^{3}}\right)^{-1} \,, \tag{24}$$

where $\mathrm{c}_{\mathrm{S}} = \sqrt{\gamma p/\rho}$ denotes the speed of sound and $\gamma$ is the adiabatic index.

## 2.2. Non-ideal terms

The system of Eqs. 1–4 contains the Ohmic and ambipolar diffusion terms, Hall term and Biermann battery term, with coefficients calculated as

$$\eta = \frac{\alpha_{\mathrm{e}}}{(en_{\mathrm{e}})^2}; \qquad \eta_{\mathrm{A}} = \frac{\xi_{\mathrm{n}}^2 \mathbf{B}^2}{\alpha_{\mathrm{n}}}; \qquad \eta_{\mathrm{H}} = \frac{|\mathbf{B}|}{en_{\mathrm{e}}}. \tag{25}$$

Here $\xi_{\mathrm{n}} = \rho_{\mathrm{n}}/\rho$ is the fraction of neutrals, $\rho_{\mathrm{n}}$ the neutral mass density, and $e$ the electron charge. The neutral and electron collisional parameters, $\alpha_{\mathrm{n}}$ and $\alpha_{\mathrm{e}}$, are defined as

$$\alpha_{\mathrm{n}} = \sum_{\beta=1}^{N} \rho_{\mathrm{e}}\nu_{\mathrm{en}_{\beta}} + \sum_{\alpha=1}^{N}\sum_{\beta=1}^{N} \rho_{\mathrm{i}_{\alpha}}\nu_{\mathrm{i}_{\alpha}\mathrm{n}_{\beta}}, \tag{26}$$

$$\alpha_{\mathrm{e}} = \sum_{\alpha=1}^{N} \rho_{\mathrm{e}}\nu_{\mathrm{ei}_{\alpha}} + \sum_{\beta=1}^{N} \rho_{\mathrm{e}}\nu_{\mathrm{en}_{\beta}}. \tag{27}$$

Here, following Khomenko *et al.* (2014a), the summation goes over $N$ neutral species (index $\mathrm{n}_{\beta}$) and $N$ singly ionized species (index $\mathrm{i}_{\alpha}$) composing the plasma. Accordingly, $\rho_{\mathrm{e}}$ is the electron mass density and $\rho_{\mathrm{i}_{\alpha}}$ the mass density of ions of species $\alpha$. Expressions for the ion-neutral ($\nu_{\mathrm{i}_{\alpha}\mathrm{n}_{\beta}}$) and electron-neutral ($\nu_{\mathrm{en}_{\beta}}$) collisional frequencies entering the collisional parameters are taken from Spitzer (1956). For the collisions between electrons and ions ($\nu_{\mathrm{ei}_\alpha}$), we use the expressions from (Braginskii, 1965):

$$\nu_{\mathrm{i}_\alpha \mathrm{n}_\beta} = n_{\mathrm{n}_\beta} \sqrt{\frac{8k_{\mathrm{B}}T}{\pi m_{\mathrm{i}_\alpha \mathrm{n}_\beta}}} \sigma_{\mathrm{in}}, \quad (28)$$

$$\nu_{\mathrm{en}_\beta} = n_{\mathrm{n}_\beta} \sqrt{\frac{8k_{\mathrm{B}}T}{\pi m_{\mathrm{en}_\beta}}} \sigma_{\mathrm{en}}, \quad (29)$$

$$\nu_{\mathrm{ei}_\alpha} = \frac{n_{\mathrm{e}} e^4 \ln \Lambda}{3\epsilon_0^2 m_{\mathrm{e}}^2} \left( \frac{m_{\mathrm{e}}}{2\pi k_{\mathrm{B}} T} \right)^{3/2}, \quad (30)$$

where $m_{\mathrm{i}_\alpha \mathrm{n}_\beta} = m_{\mathrm{i}_\alpha} m_{\mathrm{n}_\beta} / (m_{\mathrm{i}_\alpha} + m_{\mathrm{n}_\beta})$, $m_{\mathrm{en}_\beta} = m_{\mathrm{e}} m_{\mathrm{n}_\beta} / (m_{\mathrm{e}} + m_{\mathrm{n}_\beta})$ are the reduced masses, $m_{\mathrm{e}}$ the electron mass, $n_{\mathrm{n}_\beta}$ the number density of neutrals of type $\beta$, and $\epsilon_0$ the permittivity of free space. The cross sections for a weakly ionized plasma assuming elastic collisions between solid spheres are $\sigma_{\mathrm{in}} = 5 \times 10^{-19} \mathrm{m}^2$ and $\sigma_{\mathrm{en}} = 10^{-19} \mathrm{m}^2$ (Huba, 2013). The Coulomb logarithm, $\ln \Lambda$, is defined as

$$\ln \Lambda = \frac{12\pi (\epsilon_0 k_{\mathrm{B}} T)^{3/2}}{n_{\mathrm{e}}^{1/2} e^3}. \quad (31)$$

The computation of the $\eta$ coefficients, Eqs. 25, requires the knowledge of the electron number density and the number densities of different neutrals and ions composing the plasma. These are not the variables evolved in the MHD equations, Eqs. 1–4. The partial number densities in MANCHA3D are computed from the equation of state, see Sect. 2.5.

## 2.3. Radiative transfer equation

### *2.3.1. Basics*

Interactions between radiation and matter are of critical importance for the modeling of stellar atmospheres (Hubeny and Mihalas, 2014; Rutten, 2003). For an MHD model in LTE, these interactions are fully described by a net radiative energy exchange $Q_{\mathrm{R}}$ that appears as a source term in the energy equation (Eq. 3). The $Q_{\mathrm{R}}$ term is defined either based on the radiative flux $\mathbf{F}$ or on the mean radiative intensity $J$,

$$Q_i^{\mathrm{F}} = -\nabla \cdot \mathbf{F}_i, \quad (32)$$

or

$$Q_i^{\mathrm{J}} = 4\pi \varkappa_i \rho (J_i - S_i), \quad (33)$$

where $\varkappa_i$ is the absorption coefficient per unit mass, $S_i$ is the source function, and the index $i$ stands either for the radiation frequency $\nu$ or for a discrete index of a statistically representative frequency group. The flux, $F_i$, and the mean intensity, $J_i$, are computed from the known spatial and angular distribution of the specific intensity of the radiation $I_i$:

$$\mathbf{F}_i = \int_{4\pi} \mu I_i(\mu) \mathrm{d}\omega, \quad (34)$$

$$J_i = \frac{1}{4\pi} \int_{4\pi} I_i(\mu) \mathrm{d}\omega, \tag{35}$$

where $\mu = \cos\theta$ gives the direction of the ray, and the integration is done over the solid angle $\omega$. The specific intensity is, on the other hand, defined as the proportionality coefficient between the energy transported by radiation through a given area in a given direction and time interval, and at a given frequency (e.g. Eq. 2.1 of Rutten, 2003). The radiation, $I_i$, is evaluated along the ray, so that it is a function of the distance along the ray $s$ only. The definition of $I_i$ along the ray leads directly to the radiative transfer equation (RTE) in the form:

$$\frac{\mathrm{d}I_i(s)}{\varkappa_i(s)\rho(s)\mathrm{d}s} = S_i(s) - I_i(s) \tag{36}$$

where $S_i(s)$ is the source function set by the Planck function in LTE. Due to the intrinsic non-locality of the radiation field and the complexity of the opacity function ($\varkappa$), the solution of RTE presents a challenging problem even when LTE is assumed.

*2.3.2. Numerics*

In Mancha3D the RTE is solved using the short-characteristics (SC) method (Mihalas, Auer, and Mihalas, 1978; Olson and Kunasz, 1987; Kunasz and Auer, 1988). The method relies on computing the intensity contributions along each ray on short segments or characteristics using the formal solution between two adjacent points, the upwind (U) point in which the intensity is already known and the local (L) point in which it is being evaluated:

$$I(\tau_{\mathrm{L}}) = I(\tau_{\mathrm{U}})\mathrm{e}^{-\Delta\tau_{\mathrm{UL}}} + \int_{\mathrm{U}}^{\mathrm{L}} S(\tau)\,\mathrm{e}^{-(\tau_{\mathrm{L}}-\tau)}\,\mathrm{d}\tau, \tag{37}$$

where $S(\tau)$ is the source (Planck) function approximated by a polynomial on the UL segment and $\tau_{\mathrm{UL}}$ is the difference between the optical depth in the points L and U computed as:

$$\Delta\tau_{\mathrm{UL}} = \int_{\mathrm{U}}^{\mathrm{L}} \varkappa(s)\rho(s)\mathrm{d}s. \tag{38}$$

In that way, the local intensity $I(\tau_{\mathrm{L}})$ depends on all intensities upwind along the ray from the point L through the first term in Eq. 37 and on the local contributions to the radiation field between the points U and L through the second term. In the SC method, in more than one dimension, the U point lies between points of the computational grid and, therefore, the intensity in that point has to be computed by interpolation of the intensities in the adjacent points. The same applies to other values that are required at the U point, i.e. $\varkappa_{\mathrm{U}}, \rho_{\mathrm{U}}$ and $S_{\mathrm{U}}$. The required interpolation is one-dimensional if the ray is traced in 2D, and it is two-dimensional if it is traced in 3D. In the public version of Mancha3D we use the linear and bilinear interpolation formulae for these two cases, respectively. Because of the need for the interpolation in the U points, the SC method is more numerically diffusive than the long-characteristics variant of the formal solver (Peck, Criscuoli, and Rast, 2017). However, the SC method is particularly suitable for implementation in 3D geometry where parallelization is done by domain decomposition, and, therefore, it is widely accepted as the method for solving the RTE in the radiative MHD codes (Vögler, 2003; Gudiksen *et al.*, 2011) and in spectral radiative 3D codes (Ibgui *et al.*, 2013). A variant of the method with a linearly approximated source function is implemented in the publicly available version of MANCHA3D. The coefficients for numerical computation of the integral in Eq. 37 are derived in direction-independent form as in Auer and Paletou (1994). For the higher-order solution and analysis of errors, see Vitas et al. (in prep).

Equation 37 must be evaluated for any grid point in the simulation domain progressively moving along a ray from the boundary at which the ray enters the domain to the boundary at which it leaves it. As our simulation domain is decomposed to enable parallel computing, no ray is fully contained in one of the subdomains and, therefore, rays are cast from one subdomain boundary to another, and the solution has to be iterated until the relative differences at the subdomain boundaries become lower than a specified tolerance (usually the relative error of $10^{-2}$ provides sufficiently accurate solution). Since the subdomain decomposition is uniform in terms of size and shape, it takes the same number of algebraic operations, i.e. evaluations of Eq. 37, in each subdomain to reach its boundary. Once the boundary is reached, the intensities at the boundary are communicated to the next subdomain downwind in the direction of the ray propagation where these intensities become the initial ones for the next iteration step. The communications are done per direction, in order to secure that the values in the corners of the subdomain are properly updated (see enlightening Figure 4.3 of Vögler, 2003).

The global boundary condition is specified at the entering boundary for each ray. For the rays entering through the top boundary of the domain, we set that $I_i(0) = 0$, i.e. we assume that the simulation domain is not illuminated from above. For the rays entering the domain through the bottom boundary, we set that the intensity is equal to the source function, $I_i(0) = S_i(0)$, which is a very good assumption below the surface. Finally, for the rays entering through periodic vertical boundaries we assume either the same ($I_i(0) = S_i(0)$) or that the intensity is known from a previous time-step.

The angular discretization, i.e. the number and the orientation of the rays, can be defined by the user as a set of $n_\mu$ pairs $(\mu, \omega_\mu)$ where $\omega_\mu$ is the weight of each ray for the intensity quadrature over the solid angle. The distribution of the rays cannot be arbitrary. The optimal quadrature "Set A" with 3 rays per octant, proposed by Carlson (1963), is hard-coded, but it is trivial to replace it with whatever other variant (see, for example, Jaume Bestard, Štěpán, and Trujillo Bueno, 2021).

The RTE solver computes $Q_\mathrm{R}$ from three quantities: the mass density, the opacity per mass and the Planck function. The latter two must be precomputed by the user (as functions of $T$ and either $p$ or $\rho$) and stored as lookup tables. The solver is ignorant of the content of the tables, so it is the user's responsibility to provide physically correct values. These values may be monochromatic opacities, values of opacity distribution function, or opacities grouped into statistically representative bins (Nordlund, 1982). However, note that the implemented integration of the individual contributions is valid only for the cases of monochromatic opacities and opacity bins, while the integration over ODF requires a modification of the code to take into account the weights of individual ODF segments. For a detailed description of the opacity binning method, the strategies how the bins can be constructed and the intrinsic uncertainties of the method, see Perdomo García *et al.* (2023).

The two definitions of $Q_\mathrm{R}$, Eqs.32 and 33, are analytically interchangeable. However, as Bruls, Vollmöller, and Schüssler (1999) recognized, the latter is becoming numerically inaccurate in the optically thick regime and the former in the optically thin. Following the suggestion by Bruls, Vollmöller, and Schüssler (1999) (see also Vögler, 2003), we compute both solutions and $Q_i$ is evaluated as their weighted mean:

$$Q_i = \mathrm{e}^{-\tau_i/\tau_0}\, Q_i^\mathrm{J} + (1 - \mathrm{e}^{-\tau_i/\tau_0})\, Q_i^\mathrm{F} \tag{39}$$

where $\tau_i$ is optical depth in the opacity group $i$ and $\tau_0 = 0.1$. The final $Q_\mathrm{R}$ is computed as the sum over $i$, $Q_\mathrm{R} = \sum_i Q_i$.

## 2.4. Newton cooling

Alternatively, the radiative losses in Mancha3D may be computed following the Newton's cooling law:

$$Q_\mathrm{R} = -c_v \frac{T_1}{\tau_\mathrm{R}}, \tag{40}$$

where $T_1$ is the perturbation in the temperature with respect to the equilibrium value, see Sect. 2.6, $\tau_\mathrm{R}$ is the radiative relaxation time (read from a user-specified precomputed file), and $c_v$ is the specific heat at constant volume, computed assuming the equation of state of an ideal gas. This expression is valid for optically thin disturbances, for which the wavelength is much smaller than the photon mean free path. The study of propagation of acoustic waves in a radiating fluid using Newton's cooling law predicts an adiabatic propagation for waves with periods significantly shorter than $\tau_\mathrm{R}$, while in the opposite case acoustic waves propagate isothermally. However, at low enough frequencies the wavelength of the fluctuations becomes long enough so that the perturbation becomes optically thick, and the Newtonian cooling approximation is no longer valid.

## 2.5. Equation of state

The equation of state (EOS) closes the system of MHD equations coded in Mancha3D. It provides transformation functions between the various thermodynamic (TD) quantities used in the code. The primary purpose is to compute the gas pressure (required in the momentum equation) and the temperature (required for the RT module, conduction and non-ideal MHD terms) from the primary variables density and internal energy. In addition, the EOS in the code is used to compute the electron density (required for computation of the non-ideal MHD terms) and to provide inverse TD relations between $T$, $p$, $e_\mathrm{int}$ and $\rho$ that are required in some of the initial and/or boundary conditions. In practice, the EOS implementation can vary from trivial to extremely complex. In Mancha3D the EOS module supports three implementations assuming thermodynamic equilibrium.

### *2.5.1. Ideal gas*

Under the approximation of the classical ideal gas the relation between the total gas pressure $p$, the total number density of the free particles $n$ and the temperature $T$ is:

$$p = nk_\mathrm{B}T. \tag{41}$$

Or, in terms of the density:

$$p = \rho \frac{R}{\mu_{\mathrm{g}}} T \tag{42}$$

where $R$ is the universal gas constant and $\mu_{\mathrm{g}}$ is the mean molar mass of the gas. As there are no interactions between the particles in this approximation, the mean molar mass of the gas is constant with time and determined only by the chemical composition of the gas. The internal energy of the monoatomic ideal gas has only the translational component:

$$e_{\mathrm{int}} = \frac{3}{2} n k_{\mathrm{B}} T = \frac{3}{2} p \tag{43}$$

*2.5.2. Realistic EOS*

The ideal gas approximation fails in the stellar atmospheres where particles undergo dynamic interactions causing the perpetual change of their intrinsic state through molecular formation/dissociation, and atomic and molecular ionization and recombination. If the TD equilibrium is assumed (so that the distribution of the particles over the ionization stages and excitation states is given by the Saha-Boltzmann formulae and that the molecular formation is instantaneous and well described by the instantaneous chemical equilibrium, effectively an equivalent of Saha-Boltzmann), replacing the ideal EOS means to find the partial number density of all involved species of particles. This is numerically challenging and computationally expensive. The problem is even more complicated at the high densities in the deep convection zone (where processes like pressure ionization and Coulomb interaction have to be taken into account) and in the high chromosphere (where the gas and the radiation are out of TD equilibrium, so the dynamic non-equilibrium ionization/recombination has to be solved). The common approach to save on computing time in the MHD codes for the stellar atmospheres is to use precomputed lookup tables of the EOS and to interpolate them for the real-time input in the code. This approach, feasible only if the TD equilibrium is assumed, is implemented in Mancha3D.

*2.5.3. Electron pressure*

The electron pressure needs to be computed in the experiments including the non-ideal terms (ambipolar diffusion, the Biermann battery). In Mancha3D it can be either precomputed and stored in lookup EOS tables with other quantities or it can be computed on-the-fly from the pressure and the temperature. On-the-fly computations are based on the work of Vardya (1965) (initial formulation), Mihalas (1967) (implementation) and Wittmann (1974) (some corrections and algorithm described in details). These routines solve the EOS for a gas with a given set of abundances. Hydrogen is treated as neutral H, protons, negative hydrogen ion, neutral and ionized $H_2$ molecule. All other elements can be neutral or singly ionized. The system of equations then consists of one nuclei conservation equation for each species and the charge conservation equation for the electrons. These routines include an iterative solver for a system of non-linear equations and, thus, they are computationally expensive.

We note that computing the electron density this way is not fully consistent with the approximation of ideal gas because the ionization changes the number of particles and, thus, the mean molar mass, while it is constant in the ideal gas equation.

### 2.6. Split variables

The variables in the code are split into two parts, equilibrium and non-linear perturbation,

$$\begin{aligned} \rho &= \rho_0 + \rho_1; \quad & \mathbf{B} &= \mathbf{B}_0 + \mathbf{B}_1; \quad & e_{\mathrm{tot}} &= e_{\mathrm{tot},0} + e_{\mathrm{tot},1}; \\ p &= p_0 + p_1; \quad & T &= T_0 + T_1; \quad & e_{\mathrm{int}} &= e_{\mathrm{int},0} + e_{\mathrm{int},1}, \end{aligned} \tag{44}$$

where the background value is denoted with index 0 and the perturbation with index 1. Velocity has only the perturbation component, $\mathbf{v} = \mathbf{v}_1$, as it is assumed to be zero in the state of equilibrium.

The equilibrium state strictly assumes magnetohydrostatic (MHS) equilibrium in the absence of external forces ($\mathbf{v} = 0$ and $\mathbf{S} = 0$) to be fulfilled,

$$\nabla \cdot \left[ \left( p_0 + \frac{\mathbf{B_0}^2}{2\mu_0} \right) \mathbf{I} - \frac{\mathbf{B}_0 \mathbf{B}_0}{\mu_0} \right] = \rho_0 \mathbf{g}. \tag{45}$$

The set of equilibrium variables used in the initialization phase of a simulations must be consistent with this definition.

Many different kinds of equilibrium fulfill Eq. 45. For example, a force-free equilibrium or a potential magnetic field equilibrium, where the gravitationally stratified thermodynamic variables are independent from the magnetic variables, can be used to initiate a simulation with Mancha3D. Also, a trivial equilibrium where all variables with the 0 index are zero is allowed by the code. In that latter case, the full-variable equations are recovered and solved by the code, despite formally keeping the equations in their split form.

After subtracting the MHS equilibrium condition, Eq. 45, from the system of Eqs. 1–4, the following conservative system of MHD equations for non-linear perturbations of density, momentum, magnetic field and energy is solved by Mancha3D,

$$\frac{\partial \rho_1}{\partial t} + \nabla \cdot (\rho \mathbf{v}) = \left( \frac{\partial \rho_1}{\partial t} \right)_{\mathrm{diff}}, \tag{46}$$

$$\begin{aligned} \frac{\partial \rho \mathbf{v}}{\partial t} + \nabla \cdot \Bigg[ \rho \mathbf{v}\mathbf{v} + \left( p_1 + \frac{\mathbf{B}_1^2 + 2\mathbf{B}_1 \cdot \mathbf{B}_0}{2\mu_0} \right) \mathbf{I} - \\ \frac{\mathbf{B}_0 \mathbf{B}_1 + \mathbf{B}_1 \mathbf{B}_0 + \mathbf{B}_1 \mathbf{B}_1}{\mu_0} \Bigg] = \rho_1 \mathbf{g} + \mathbf{S}(t) + \left( \frac{\partial \rho \mathbf{v}}{\partial t} \right)_{\mathrm{diff}}, \end{aligned} \tag{47}$$

$$\begin{aligned} \frac{\partial e_{\mathrm{tot},1}}{\partial t} + \nabla \cdot \Bigg[ \mathbf{v} \left( e_{\mathrm{tot}} + p + \frac{\mathbf{B}^2}{2\mu_0} \right) - \frac{\mathbf{B}(\mathbf{v} \cdot B)}{\mu_0} + \frac{(\eta_{\mathrm{A}} + \eta) \mathbf{J}_\perp \times \mathbf{B}}{\mu_0} - \\ \frac{\nabla p_{\mathrm{e}} \times \mathbf{B}}{e n_{\mathrm{e}} \mu_0} + \mathbf{q} \Bigg] = (\rho \mathbf{g} + \mathbf{S}(t)) \cdot \mathbf{v} + Q_{\mathrm{R}} + \left( \frac{\partial e_{\mathrm{tot},1}}{\partial t} \right)_{\mathrm{diff}}, \end{aligned} \tag{48}$$

$$\frac{\partial \mathbf{B}_1}{\partial t} = \nabla \times \left[ \mathbf{v} \times \mathbf{B} - \eta \mathbf{J} - \eta_{\mathrm{A}} \mathbf{J}_\perp + \frac{\nabla p_{\mathrm{e}}}{e n_{\mathrm{e}}} - \eta_{\mathrm{H}} \frac{(\mathbf{J} \times \mathbf{B})}{|\mathbf{B}|} \right] + \left( \frac{\partial \mathbf{B}_1}{\partial t} \right)_{\mathrm{diff}}. \tag{49}$$

The use of equations for perturbations instead of the full variables has several advantages for different classes of simulations. First, the terms describing the static model and those for perturbations can vary by orders of magnitude. Thus, by excluding equilibrium terms we avoid important numerical precision problems. This method also avoids the numerical diffusion acting on equilibrium quantities. Secondly, the boundary conditions are easier to implement for the equations for perturbations (see Section 3.7). Finally, if the equilibrium is computed numerically, a full-variable simulation may still evolve despite the zero perturbation due to the numerical errors in the derivatives. This can cause non-desirable spurious change of equilibrium, compromising the physical simulation. On the contrary, the simulation with a non-zero MHS equilibrium and strictly zero perturbation will produce strictly zero output in Mancha3D.

The numerical treatment of the full and split variables is slightly different in the code due to numerical diffusivities, see Section 3.3. The artificial diffusion coefficients are always computed with respect to the perturbation part only. If a non-zero MHS equilibrium (split variables) is used, it eliminates the influence of a stratified atmosphere on the perturbation evolution. Moreover, it often allows using smaller diffusion coefficients which improves numerical resolution of a problem.

While the system Eqs. 46–49 is solved by default, the code can be configured to solve fully linear equations, where all terms containing products of two perturbed variables are explicitly omitted. The linear configuration has been mostly used in simulations of waves in the solar interior and lower atmosphere (Felipe, Crouch, and Birch, 2013; Felipe *et al.*, 2016; Felipe, Braun, and Birch, 2017).

## 3. Numerical techniques

In this section we discuss the time integration methods, available spatial discretization schemes and other numerical features related to the code stability and robustness. As the code is multi-purpose, it does not have any predefined boundary condition, and it is the user's responsibility to treat the boundaries according to the particular setup.

### 3.1. Time integration schemes

For efficient usage of the computational resources several numerical schemes are implemented in Mancha3D. By default, the governing equations 46–49 are evolved by means of a memory-saving variant of the explicit RK scheme [3]. Two more integration schemes can be used to deal with the non-ideal terms in the energy and induction equations, Eqs. 48 and 49. The Battery term is usually small in solar atmosphere, and does not require modifications of the MHD integration time step. On the other hand, under certain conditions, numerical treatment of the ambipolar, Ohmic and Hall terms can become troublesome. The ambipolar term becomes dominant from the middle chromosphere upwards, and also in regions with strong magnetic field (Khomenko *et al.*, 2014a). Due to its parabolic nature, it can strongly limit the time step. The Hall term can become dominant in the middle photosphere in regions with weaker fields. It introduces dispersion, produces whistler waves and, if not treated properly, it can lead to numerical instabilities (see e.g., Tóth, Ma, and Gombosi, 2008). To handle those terms, Mancha3D can use the STS (for ambipolar and Ohmic term) and HDS (for the Hall term) schemes (O'Sullivan and Downes, 2006, 2007) to overcome time step limitations. Both techniques can be used together by applying the Strang operator splitting (Strang, 1968); the implementation of these schemes in Mancha3D (González-Morales *et al.*, 2018) is briefly summarized below.

[3]This scheme does not fulfill all the mathematical conditions to belong to the RK family, so rigorously speaking it should be called as "explicit multistage scheme". Still, for historical reasons we keep calling it as "the RK scheme" in the paper.

### *3.1.1. Memory-saving time integration scheme*

The system of Eqs. 46–49, written in conservative form, can be represented as:

$$\frac{\partial \mathbf{u}}{\partial t} = \mathcal{R}(\mathbf{u}) = -\nabla \mathbf{F}(\mathbf{u}) + \mathbf{S}(\mathbf{u}), \tag{50}$$

where the operator $\mathcal{R}(\mathbf{u})$ is the sum of the divergence of fluxes $-\nabla\mathbf{F}(\mathbf{u})$ and of the source terms, $\mathbf{S}(\mathbf{u})$. The vector $\mathbf{u}$ stands for the primary variables: $\mathbf{u} = [\rho_1, \rho\mathbf{v}, e_{\mathrm{tot},1}, \mathbf{B_1}](\mathbf{r}, t)$.

The Eqs. 46–49 are advanced in time using a variant of the explicit RK scheme that is written in a compact form as:

$$\mathbf{u}^{(k)} = \mathbf{u}^{(n)} + \alpha_k \Delta t \mathcal{R}\left(\mathbf{u}^{(k-1)}\right), \quad k = 1, \ldots, m, \tag{51}$$

$$\mathbf{u}^{(n+1)} = \mathbf{u}^{(k=m)}, \tag{52}$$

where $\mathbf{u}^{(n+1)}$ corresponds to the solution at $t_{n+1} = t_n + \Delta t$, and $\mathbf{u}^{(k)}$ is an intermediate solution at substep $k$, the coefficients $\alpha_k$ are computed using the expression,

$$\alpha_k = \frac{1}{m+1-k} \tag{53}$$

where $m$ is the order of the scheme; the same method was adopted in the code MURaM (Vögler, 2003; Vögler *et al.*, 2005). The scheme is memory efficient as it is not necessary to keep the solutions from intermediate stages. The 2nd order, two-stage scheme recovers the midpoint method, however the higher order schemes deviate from the standard ones, lowering the actual accuracy order.

The multistage RK scheme is usually described by a Butcher tableau, formed by several coefficients (Butcher, 1987). If one requires the method to have a certain order, the coefficients must fulfill several conditions derived from a Taylor expansion. The coefficients (53) always match the 1st and 2nd order conditions (see Sect. 31 in Butcher, 1987), while for higher order not all the conditions are fulfilled. For example, for the 3rd order scheme the coefficients (53) satisfy five of the six conditions (one condition of the 3rd order is not fulfilled); for the 4th order scheme only seven restricting conditions out of eleven are fulfilled (three conditions of the 4th order and one condition of the 3rd order are not fulfilled). Hence this scheme with four or more stages does not have the desired order of accuracy reaching only the 2nd one; this makes the 3rd order scheme computationally more efficient as compared to the higher order schemes of this stencil.

We run a series of tests to check the accuracy order of the scheme. For a smooth solution the accuracy order can be estimated as,

$$A = \log_2 \left( \frac{u_{\Delta t} - u_{\Delta t/2}}{u_{\Delta t/2} - u_{\Delta t/4}} \right), \tag{54}$$

where $u_{\Delta t}$, $u_{\Delta t/2}$, and $u_{\Delta t/4}$ are the solutions computed with the corresponding time steps indicated by the indices. Using this expression we obtain $A$ between 2 and 2.1 for the 2nd order scheme and $A$ between 2.5 and 2.8 for the 3rd order scheme, depending on the reference time step. The 4th order scheme produces $A \approx 2.1$, which is in line with the discussion above. For this reason, by default, MANCHA3D uses the 3rd order scheme to solve the non-ideal MHD equations, which is explicitly written as follows:

$$\begin{aligned} \mathbf{u}^{(1/3)} &= \mathbf{u}^{(0)} + \frac{\Delta t}{3}\mathcal{R}(\mathbf{u}^{(0)}), \\ \mathbf{u}^{(1/2)} &= \mathbf{u}^{(0)} + \frac{\Delta t}{2}\mathcal{R}(\mathbf{u}^{(1/3)}), \\ \mathbf{u}^{(1)} &= \mathbf{u}^{(0)} + \Delta t\mathcal{R}(\mathbf{u}^{(1/2)}). \end{aligned} \tag{55}$$

Still, user is free to use a different scheme order.

### *3.1.2. STS operator*

The STS scheme is used to speedup simulations which involve the Ohmic and ambipolar terms. The Ohmic term is usually small in solar conditions, but it is included in the STS operator for some particular applications. These terms appear only in the energy and magnetic field equations, so the Eq. 50 is written as

$$\frac{\partial \mathbf{w}}{\partial t} = \mathcal{S}(\mathbf{w}) = L^{\mathrm{Ohm}}(\mathbf{w}) + L^{\mathrm{Ambi}}(\mathbf{w}), \tag{56}$$

where the vector $\mathbf{w}$ corresponds to either $\mathbf{w} = [\mathbf{B_1}, e_{\mathrm{tot}}](\mathbf{r}, t)$ or $\mathbf{w} = [\mathbf{B_1}, e_{\mathrm{int}}](\mathbf{r}, t)$. The components of the operator $\mathcal{S}(\mathbf{w})$ are defined as,

$$\begin{aligned} \mathcal{S}(\mathbf{B_1}) &= -\nabla \times [\eta \mathbf{J} + \eta_{\mathrm{A}} \mathbf{J}_\perp], \\ \mathcal{S}(e_{\mathrm{tot}}) &= -\nabla \cdot \left[ \frac{(\eta + \eta_{\mathrm{A}})\mathbf{J}_\perp \times \mathbf{B}}{\mu_0} \right], \\ \mathcal{S}(e_{\mathrm{int}}) &= \eta J^2 + \eta_{\mathrm{A}} {J_\perp}^2. \end{aligned} \tag{57}$$

For the STS scheme the stability is imposed at the end of a bigger step called super-step, $\Delta t_{\mathrm{STS}}$. The $\Delta t_{\mathrm{STS}}$ is calculated as a sum of sub-steps $\tau_j$, obtained using the modified Chebyshev polynomials (Alexiades, Amiez, and Gremaud, 1996).

$$\tau_j = \Delta t_{\mathrm{diff}} \left[ (\nu - 1) \cos\left( \frac{2j - 1}{N_{\mathrm{STS}}} \frac{\pi}{2} \right) + 1 + \nu \right]^{-1}; \qquad \Delta t_{\mathrm{STS}} = \sum_{j=1}^{N_{\mathrm{STS}}} \tau_j, \tag{58}$$

where the minimum time step $\Delta t_{\mathrm{diff}}$ is mainly determined by the ambipolar term ($\Delta t_{\mathrm{Ambi}} = \min(\Delta x^2, \Delta y^2, \Delta z^2)/\eta_{\mathrm{A}}$)). The parameter $N_{\mathrm{STS}}$ is the number of sub-steps, and $\nu$ is a damping parameter associated with the Chebyshev polynomials. The values of the pair of the parameters $(N_{\rm STS}, \nu)$ controls the stability, accuracy and speed of the method. By selecting $\nu$ close to unity the super-step becomes smaller and the method is more accurate, but the gain from applying the STS scheme decreases. By selecting $\nu$ close to zero the system may become unstable at some point, but the benefit from applying the STS is larger (see González-Morales *et al.*, 2018, for details).

The time update by the STS operator can be written as

$$\mathbf{w}(\mathbf{r}, t_{n+1}) = \mathbf{w}(\mathbf{r}, t_n) + \sum_{j=1}^{N_{\rm STS}} \tau_j \frac{\partial \mathbf{w}(\mathbf{r}, t)}{\partial t}\Big|_{t_n+\sum_{k=1}^{j-1}\tau_k}, \tag{59}$$

where $t_{n+1} = t_n + \Delta t_{\rm STS}$. The STS scheme is a first order accurate. In Mancha3D, we consider the STS scheme as an "Eulerian" first-order step of our multi-step RK scheme, and we reach, for example, the 3rd order accuracy by applying three calls to the STS scheme. Thus, the complete temporal scheme for the STS operator can be written similarly to Eqs. (51) but, in this case, using the operator $\mathcal{S}(\mathbf{w})$ and its corresponding time step $\Delta t_{\rm STS}$:

$$\begin{aligned} \mathbf{w}^{(k)} &= \mathbf{w}^{(n)} + \alpha_k \Delta t_{\rm STS} \mathcal{S}\left(\mathbf{w}^{(k-1)}\right), \qquad k = 1, ..., m, \\ \mathbf{w}^{(n+1)} &= \mathbf{w}^{(m)}, \end{aligned} \tag{60}$$

with $\alpha_k$ given by Eq. (53).

### 3.1.3. HDS operator

The HDS operator only solves the Hall term in the induction equation. It is designed by O'Sullivan and Downes (2006) to overcome the problems originated by a skew-symmetric Hall term dominated system, and it can be written in conservative form as

$$\frac{\partial \mathbf{b}}{\partial t} = \mathcal{H}(\mathbf{b}) \tag{61}$$

where the vector $\mathbf{b}$ is the conserved variable $\mathbf{B_1}(\mathbf{r}, t)$ and $\mathcal{H}(\mathbf{b})$ is the Hall term operator,

$$\mathcal{H}(\mathbf{b}) = -\nabla \times \left[\eta_{\rm H} \frac{(\mathbf{J} \times \mathbf{B})}{|B|}\right]. \tag{62}$$

The term $\mathcal{H}(\mathbf{b})$ is treated as a usual MHD term, except that the update of the magnetic field components is done using all the information available at the moment. The following steps are preformed in the HDS scheme,

$$\begin{aligned} B_{1x}^{(k)} &= B_{1x}^{(n)} + \alpha_k \Delta t_{\rm Hall} \mathcal{H}\left(B_{1x}^{(k-1)}, B_{1y}^{(k-1)}, B_{1z}^{(k-1)}\right), \\ B_{1y}^{(k)} &= B_{1y}^{(n)} + \alpha_k \Delta t_{\rm Hall} \mathcal{H}\left(B_{1x}^{(k)}, B_{1y}^{(k-1)}, B_{1z}^{(k-1)}\right), \\ B_{1z}^{(k)} &= B_{1z}^{(n)} + \alpha_k \Delta t_{\rm Hall} \mathcal{H}\left(B_{1x}^{(k)}, B_{1y}^{(k)}, B_{1z}^{(k-1)}\right), \; k = 1, \ldots, m, \end{aligned} \tag{63}$$

$$\mathbf{B_1}^{(n+1)} = \mathbf{B_1}^{(m)}. \tag{64}$$

where $\Delta t_{\rm Hall}$ is the time step imposed by the Hall term, $\Delta t_{\rm Hall} = 2/\sqrt{27}\,\min(\Delta x^2, \Delta y^2, \Delta z^2)/\eta_{\rm H}$ (O'Sullivan and Downes, 2007). The HDS operator advances in time by repeating the calculation of the Equations (63) $N_{\rm HDS}$ times, so that $\Delta t_{\rm HDS} = N_{\rm HDS}\Delta t_{\rm Hall}$. If the HDS scheme is used together with the STS one, their global time step is set to be equal, $\Delta t_{\rm STS} = \Delta t_{\rm HDS}$.

### 3.2. Spatial discretization

The computational domain is discretized with three-dimensional Cartesian grid which is constant in time. By default, the code uses uniform grid which can have different grid spacing in different directions. It perfectly fits performing realistic simulations of the solar and stellar atmospheres, as modeling of turbulent convective flows requires fine uniform grid in all directions. However, other tasks, like wave propagation can have very elongated computational domain in one direction. To handle such problems more efficiently a non-uniform grid in z-direction is implemented, allowing for a finer resolution where it is needed and a coarser grid in the regions where plasma is more homogeneous.

#### *3.2.1. Uniform grid*

For the case of uniform grid two different schemes for spatial derivatives are implemented:
(1) a central difference, 4th-order accurate scheme using five-point stencil. In this case, the first derivative in the direction $s = \{x, y, z\}$ is computed as,

$$\left(\frac{\partial \mathbf{F}(\mathbf{u})}{\partial s}\right)_i = \frac{1}{12\Delta s}\Big(8\mathbf{F}(\mathbf{u})_{i+1} - 8\mathbf{F}(\mathbf{u})_{i-1} - \mathbf{F}(\mathbf{u})_{i+2} + \mathbf{F}(\mathbf{u})_{i-2}\Big); \tag{65}$$

(2) a 6th-order accurate scheme with ten-point stencil, used in the Stagger-code (Magic *et al.*, 2013; Nordlund and Galsgaard, 1995). In this case the derivatives are computed in two steps, using interpolation to a half-grid point:

$$\begin{aligned} \mathbf{F}(\mathbf{u})_{i+1/2} &= \mathrm{a}_1\Big(\mathbf{F}(\mathbf{u})_i + \mathbf{F}(\mathbf{u})_{i+1}\Big) + \\ &\quad \mathrm{b}_1\Big(\mathbf{F}(\mathbf{u})_{i-1} + \mathbf{F}(\mathbf{u})_{i+2}\Big) + \\ &\quad \mathrm{c}_1\Big(\mathbf{F}(\mathbf{u})_{i-2} + \mathbf{F}(\mathbf{u})_{i+3}\Big), \end{aligned} \tag{66}$$

and then taking derivative,

$$\begin{aligned} \left(\frac{\partial \mathbf{F}(\mathbf{u})}{\partial s}\right)_i &= \frac{\mathrm{a}_2}{\Delta s}\Big(\mathbf{F}(\mathbf{u})_{i-1/2} - \mathbf{F}(\mathbf{u})_{i+1/2}\Big) + \\ &\quad \frac{\mathrm{b}_2}{\Delta s}\Big(\mathbf{F}(\mathbf{u})_{i-3/2} - \mathbf{F}(\mathbf{u})_{i+3/2}\Big) + \\ &\quad \frac{\mathrm{c}_2}{\Delta s}\Big(\mathbf{F}(\mathbf{u})_{i+5/2} - \mathbf{F}(\mathbf{u})_{i+5/2}\Big), \end{aligned} \tag{67}$$

where the coefficients are:

$$a_1 = \frac{150}{256}; \quad b_1 = -\frac{25}{256}; \quad c_1 = \frac{3}{256}; \tag{68}$$

$$a_2 = -\frac{2250}{1920}; \quad b_2 = \frac{125}{1920}; \quad c_2 = -\frac{9}{1920}. \tag{69}$$

It should be mentioned that the second derivatives are computed by taking the first derivative twice for both uniform and non-uniform grids.

### *3.2.2. Non-uniform grid*

A non-uniform grid can be constructed in many different ways, depending on particular setup. As MANCHA3D is a multipurpose code, building an efficient non-uniform grid is left to the user. Below, we consider an example of modeling acoustic wave propagating from the photosphere to the corona. In Figure 1 we plot a temperature distribution in the solar atmosphere together with the vertical grid spacing. The temperature profile is taken from the VALC model (Vernazza, Avrett, and Loeser, 1981) and is expanded to corona with a constant temperature of $10^6$ K. The grid is defined to have the highest resolution around the transition region where the thermodynamic quantities exhibit the strongest gradients. On the other hand, in the case of non-uniform grid the local truncation error is affected by the grid stretching factor ($\Delta z_{i+1}/\Delta z_i$), it should not deviate from unity by more than 20% (Fletcher, 1988; Jianchun, Pope, and Sepehrnoori, 1995), where unity stands for uniform grid. Therefore it is important to have a smooth and gradual variation of the grid spacing similar to the one shown in the inset of Figure 1, where the grid factor is less than about 10%, i.e. $0.9 < \Delta z_{i+1}/\Delta z_i < 1.1$

The derivatives in z-direction are computed using a central-like scheme of 4th or 6th order; for example, the 1st derivative of the 4th order accuracy is computed with five-point stencil as,

$$\left(\frac{\partial \mathbf{F}(\mathbf{u})}{\partial s}\right)_i = \sum_{j=i-2}^{j=i+2} a_j \mathbf{F}(\mathbf{u})_j, \tag{70}$$

where coefficients $a_j$ are computed for particular non-uniform grid using Taylor expansion. As the grid is constant in time, they are computed only once at the initialization stage, therefore taking derivatives in the case of non-uniform grid does not have additional computational costs.

The benefit of using a non-uniform grid is twofold. First, it allows using smaller number of grid points as compared to the corresponding uniform grid covering the same computational domain. Second, depending on the particular setup, the computational time step can be noticeably larger than the one from the uniform grid. For example, if we consider a non-magnetic subsonic flow, the time step is limited by the sound speed, $c_S$, which varies along z-direction. In case of the uniform grid the time step is proportional to $\Delta t \propto \Delta z/c_{S,\max}$, where $c_{S,\max}$ corresponds to the highest sound speed from the hot corona region, and it limits the time step in the whole domain. In the case of the non-uniform grid, the grid spacing also varies and the global time step is computed from the local values,

$$\Delta t \propto \left[\frac{\Delta z_i}{c_{\mathrm{S},i}}\right]_{\min}, \tag{71}$$

where the index $i$ runs over all the grid points. Hence, having a larger grid spacing in the coronal region removes the limitation set by high sound speed in the case of the uniform grid. Still, one should keep in mind that the efficiency of the fixed non-uniform grid strongly depends on the simulation setup and, in particular, how accurately one can predict where the refinement of the grid is needed.

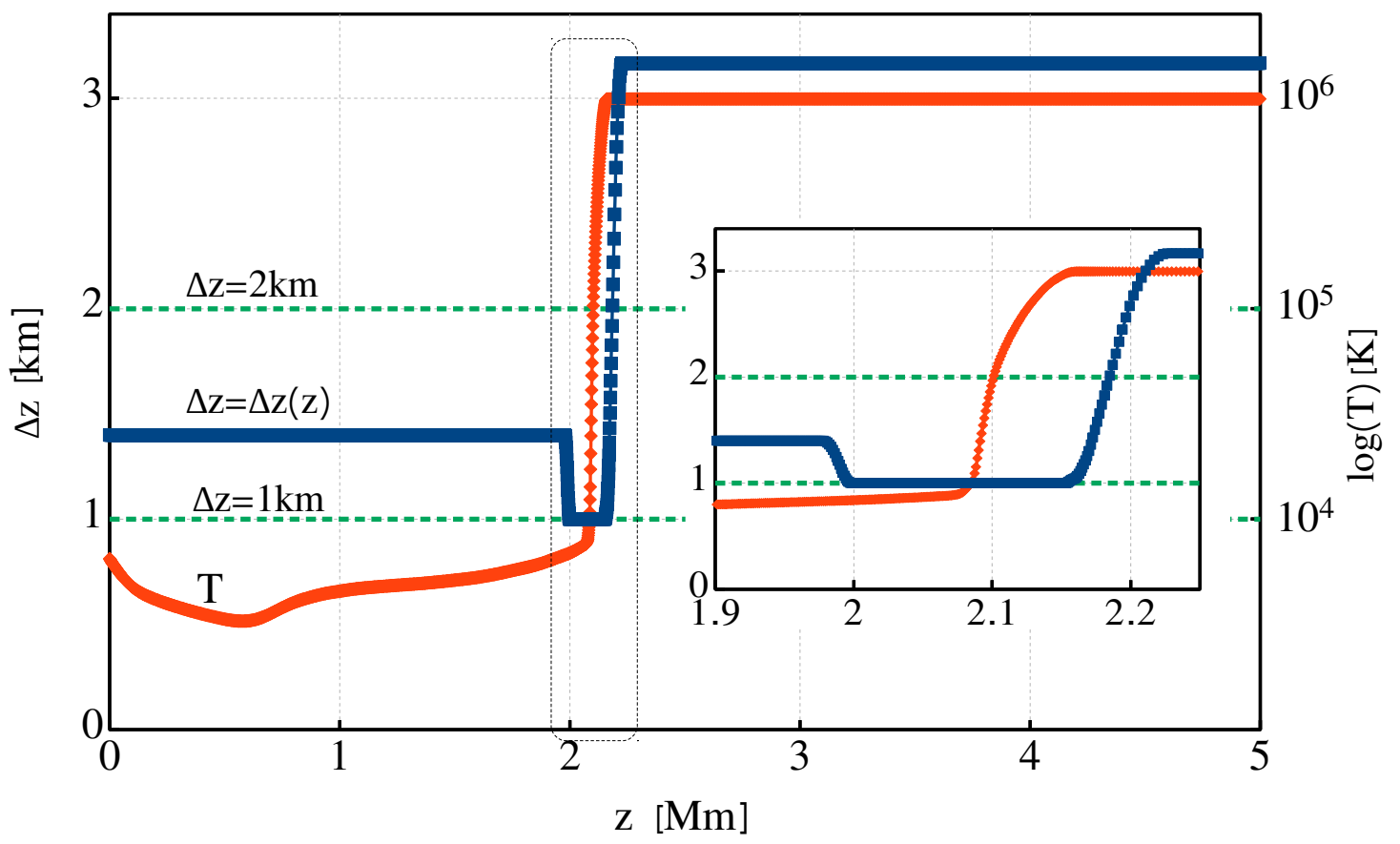

**Figure 1.** Non-uniform grid spacing (blue line) together with the temperature profile (red line) in a logarithmic scale used in the acoustic wave test. Two uniform grid spacing are plotted as dashed lines for comparison; the $\Delta z = 2$ km grid size covers the domain with the same number of grid points, N = 2500; the $\Delta z = 1$ km is the minimum grid spacing of the non-uniform grid. The inset shows the gradual variation of the non-uniform grid spacing around the transition region.

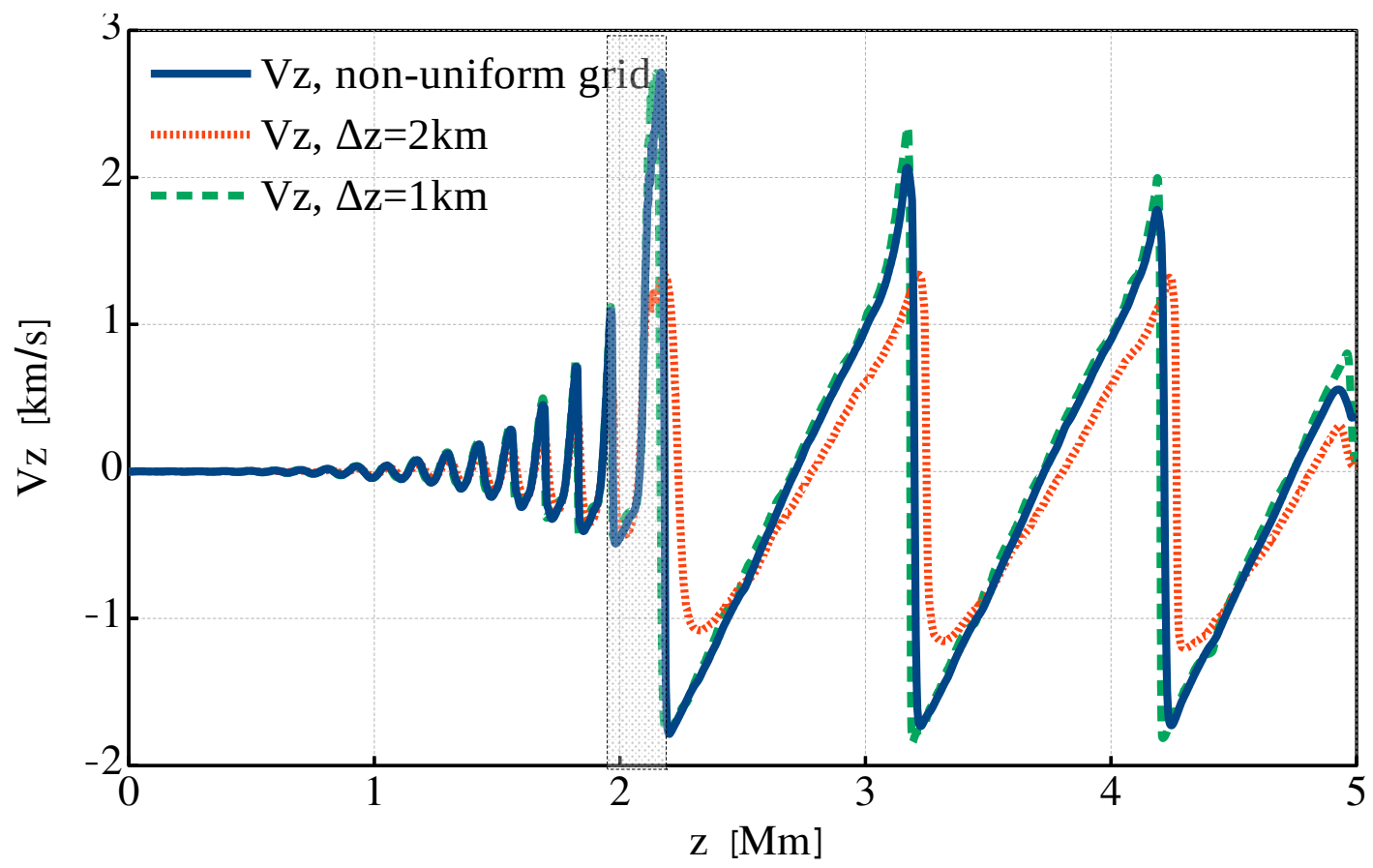

**Figure 2.** Vertical velocity profiles at time moment $t = 398$ s for the three grids shown in Figure 1; the shaded region denotes the location of the transition region with the finest resolution in the non-uniform grid case.

**Table 1.** Comparison between the three runs of the acoustic wave test. The time step is determined by advection and its Courant–Friedrichs–Lewy condition, below the averaged value is shown.

| Parameter | Run1 | Run2 | Run3 |
|---|---|---|---|
| Grid spacing [km] | 1 | 2 | 1-3.1 |
| Number of grid points | 5000 | 2500 | 2500 |
| Time step [s] | 0.0035 | 0.0068 | 0.0076 |
| Number of iterations | 115786 | 58509 | 52906 |
| Run wall-clock time [s] | 324 | 126 | 114 |

To demonstrate the advantages of the non-uniform grid we performed a test simulation of an acoustic wave propagation in a 1D atmosphere, shown in Figure 1. The atmosphere spans over 5 Mm from the photosphere to the corona. The wave is triggered at the bottom boundary by an analytical solution of the vertical velocity $v_z$, density $\rho_1$, pressure $p_1$ (together with consistent perturbations in the remaining thermodynamic quantities), with a period of 15 sec and a starting amplitude of 1 m s$^{-1}$, see Appendix A4 in Felipe, Khomenko, and Collados (2010). The PML (see Sect. 3.7) at the top boundary damps all the perturbations, preventing possible reflections.

We performed three runs with different grids, two uniform, one with 5000 grid points and $\Delta z = 1$ km and another with 2500 points and $\Delta z = 2$ km and one non-uniform, with the grid spacing shown in Figure 1. In Figure 2 we plot the velocity profiles computed for the three grids, together with the location of the transition region. Qualitatively all three profiles look similar and before reaching the transition region the difference between them is negligible. After the wave passes through transition region the velocity profile computed with the non-uniform grid is much closer to the one obtained with the finest uniform grid (solid blue and dashed green lines) than the profile computed with the same number of points, but uniformly distributed. It is important to notice that above the transition region, the non-uniform grid spacing is considerably larger than both uniform grid spacing, but velocity evolution is still much closer to the run with the finest uniform grid than the one with the coarser grid.

The computational details of these three runs are summarised in Table 1. From this table we see that the simulation with the non-uniform grid runs slightly faster as compared to the one with the coarser uniform grid and the same number of grid points. That is due to the local computation of the time step, as described above. The "Run1" with the fine uniform grid run takes $\sim 2.8$ more time due to both larger amount of grid points and smaller time step which requires to take more time steps to reach $t = 400$ s. For all three runs we used same number of CPUs, which means twice the load for the first run with the finest uniform grid.

### 3.3. Artificial diffusion

In astrophysical plasmas, both the hydrodynamic and the magnetic Reynolds number (associated to the Ohmic diffusion) usually have very large values, making the characteristic lengths in which the viscosity and diffusivity act too small to be resolved. To prevent the exponential grow of numerical noise at these small scales, Mancha3D uses artificial equivalents of the physical viscosity, magnetic diffusivity and thermal conduction, as well as a completely artificial diffusion term in the continuity equation. This approach resembles the one described in Stein and Nordlund (1998); Caunt and Korpi (2001); Vögler *et al.* (2005). We generically refer to all these terms as "artificial diffusivities".

The diffusion term in the continuity equation, Eq. 46, is defined as,

$$\left(\frac{\partial \rho_1}{\partial t}\right)_{\rm diff} = \sum_i \frac{\partial}{\partial x_i}\left[\nu_i(\rho_1)\frac{\partial \rho_1}{\partial x_i}\right], \tag{72}$$

where $\nu_i$ is the diffusion coefficient, computed as explained in Sect. 3.4. The index $i$ counts the three Cartesian directions. Note that in general the operator applies to the non-linear density perturbation, $\rho_1$. However, in the zero-equilibrium case it is equivalent to applying it to the full variable since $\rho = \rho_1$.

The diffusion term in the equation of motion, Eq. 47, is defined as,

$$\left(\frac{\partial \rho \mathbf{v}}{\partial t}\right)_{\rm diff} = \nabla \cdot \boldsymbol{\tau}, \tag{73}$$

where $\boldsymbol{\tau}$ is viscous stress tensor with components:

$$\tau_{ij} = \frac{1}{2}\rho\left(\nu_j(v_i)\frac{\partial v_i}{\partial x_j} + \nu_i(v_j)\frac{\partial v_j}{\partial x_i}\right). \tag{74}$$

In the induction equation, Eq. 49, the diffusion term is,

$$\left(\frac{\partial \mathbf{B_1}}{\partial t}\right)_{\rm diff} = -\nabla \times \boldsymbol{\varepsilon}, \tag{75}$$

where $\boldsymbol{\varepsilon}$ plays the role of an equivalent electric field vector with three components ($x$, $y$, $z$):

$$\begin{aligned}
\varepsilon_x &= \left(\nu_y(B_{1z})\frac{\partial B_{1z}}{\partial y} - \nu_z(B_{1y})\frac{\partial B_{1y}}{\partial z}\right),\\
\varepsilon_y &= \left(\nu_z(B_{1x})\frac{\partial B_{1x}}{\partial z} - \nu_x(B_{1z})\frac{\partial B_{1z}}{\partial x}\right),\\
\varepsilon_z &= \left(\nu_x(B_{1y})\frac{\partial B_{1y}}{\partial x} - \nu_y(B_{1x})\frac{\partial B_{1x}}{\partial y}\right).
\end{aligned} \tag{76}$$

Here, again, if a non-zero MHS equilibrium is used, the operator defined by Eq. 75 applies to the magnetic field perturbation $\mathbf{B_1}$, not to the full vector $\mathbf{B_0} + \mathbf{B_1}$, while for the zero-equilibrium case it applies to the full variable, $\mathbf{B} = \mathbf{B_1}$.

The artificial diffusivity term in the total energy equation is composed of three components: viscous and Ohmic heating terms, and artificial heat conduction (Vögler *et al.*, 2005),

$$\left(\frac{\partial e_{\mathrm{tot},1}}{\partial t}\right)_{\mathrm{cond}} = \sum_i \frac{\partial}{\partial x_i}\left(\rho \nu_i(T_1)\frac{\partial c_p T_1}{\partial x_i}\right), \tag{77}$$

$$\left(\frac{\partial e_{\mathrm{tot},1}}{\partial t}\right)_{\mathrm{Ohm}} = \nabla\cdot(\mathbf{B}\times\boldsymbol{\varepsilon}), \tag{78}$$

$$\left(\frac{\partial e_{\mathrm{tot},1}}{\partial t}\right)_{\mathrm{visc}} = \nabla\cdot(\mathbf{v}\cdot\boldsymbol{\tau}). \tag{79}$$

where $c_p$ is the specific heat at constant pressure. In the internal energy equation, the artificial conductivity term remains the same, while the other two are modified accordingly,

$$\left(\frac{\partial e_{\mathrm{int}}}{\partial t}\right)_{\mathrm{Ohm}} = \nabla\cdot(\boldsymbol{\varepsilon}\cdot\mathbf{J}), \tag{80}$$

$$\left(\frac{\partial e_{\mathrm{int}}}{\partial t}\right)_{\mathrm{visc}} = \boldsymbol{\tau} : \nabla\mathbf{v}, \tag{81}$$

where : stands for tensor double contraction.

### 3.4. Artificial diffusion coefficients

There are three contributions to the artificial diffusivity coefficients $\nu$ in Mancha3D. For a quantity $\mathbf{u}$ (scalar or vector) and direction $i$, the artificial diffusivity coefficient can be written as:

$$\nu_i(\mathbf{u}) = \nu_i^{\mathrm{const}}(\mathbf{u}) + \nu_i^{\mathrm{hyper}}(\mathbf{u}) + \nu_i^{\mathrm{shock}}(\mathbf{u}). \tag{82}$$

The first term $\nu^{\mathrm{const}}$ stands for the part constant in time,

$$\nu_i^{\mathrm{const}}(\mathbf{u}) = c^{\mathrm{const}}(\mathbf{u})\,(c_{\mathrm{S0}} + v_{\mathrm{A0}})\Delta x_i F^{\mathrm{const}}(x,y,z), \tag{83}$$

where $F^{\mathrm{const}}$ is a user-defined profile that accounts for the spatial variation of the constant diffusivity. The purpose of this term is to enhance the constant diffusion in a specific region, for example close to a domain boundary, while keeping it low elsewhere. The term $(c_{\mathrm{S0}} + v_{\mathrm{A0}})$ is the sum of the sound speed and the Alfvén speed computed from the initial MHS atmosphere. The coefficients $c^{\mathrm{const}}(\mathbf{u})$ are the amplitudes of the diffusivity of different primary variables $\mathbf{u} = [\rho_1, \rho\mathbf{v}, e_1, \mathbf{B_1}]$.
The variable (hyper) diffusivity term $\nu^{\mathrm{hyper}}$ is defined as:

$$\nu_i^{\mathrm{hyper}}(\mathbf{u}) = c^{\mathrm{hyper}}(\mathbf{u})\,(v + c_{\mathrm{S}} + v_{\mathrm{A}})H_i(\mathbf{u})\Delta x_i\, F^{\mathrm{hyper}}(x,y,z). \tag{84}$$

Similar to the constant counterpart, the coefficients $c^{\mathrm{hyper}}(\mathbf{u})$ are the diffusivity amplitudes of the various variables. The term $(v + c_{\mathrm{S}} + v_{\mathrm{A}})$ is computed using the local flow, and the sound and Alfvén velocities. The hyper-diffusion core term $H_i(\mathbf{u})$ is computed as,

$$H_i(\mathbf{u}) = \frac{\max_3 |3(\mathbf{u}_{h+1} - \mathbf{u}_h) - (\mathbf{u}_{h+2} - \mathbf{u}_{h-1})|}{\max_3 |\mathbf{u}_{h+1} - \mathbf{u}_h|}, \tag{85}$$

where $\max_3$ denotes the maximum over 3 adjacent points (Vögler, 2003). This ratio is defined as a mask, trimmed between 0 and 1, that takes large values at places where small-scale variations with large amplitudes are present, but keeping low values elsewhere. The function $F^{\text{hyper}}$ is a user-specified profile, allowing to enhance the amplitude of the hyper diffusion in specific predefined regions, similarly to how $F^{\text{const}}$ modifies the amplitude of the constant diffusivity.

The shock diffusivity term $\nu^{\text{shock}}$ takes high values in the regions where there are strong gradients with sudden variations in the velocity between nearby points. It is proportional to the absolute value of the divergence of the velocity only in those locations where there are converging flows, being zero in the rest of the domain (Vögler, 2003):

$$\begin{aligned} \nu_i^{\text{shock}}(\mathbf{u}) &= c^{\text{shock}}(\mathbf{u})(\Delta x_i)^2 |\nabla \cdot \mathbf{v}|, \qquad \nabla \cdot \mathbf{v} < 0, \\ \nu_i^{\text{shock}}(\mathbf{u}) &= 0, \qquad \nabla \cdot \mathbf{v} \geq 0. \end{aligned} \tag{86}$$

The parameter $c^{\text{shock}}(\mathbf{u})$ is the amplitude of the shock diffusivity that can be set independently for each of the variables $\mathbf{u}$.

### 3.5. Filtering

In some types of simulations, for example those corresponding to wave propagation, a high diffusion is not desirable since it modifies the wave amplitudes. At the same time, a low diffusion can not always prevent the development of high frequency noise. In such cases, Mancha3D can perform an additional filtering of small wavelengths, which can be applied with a user-defined frequency. We use the filtering function defined in (Parchevsky and Kosovichev, 2007) ,

$$u_{\text{filt}} = u(x) - \sum_{m=-3}^{3} \mathrm{d}_m u(x + m\Delta x), \tag{87}$$

where $u$ is the variable before filtering and $u_{\text{filt}}$ is the one after filtering. The filter can be applied in each of the three spatial directions independently. The coefficients $\mathrm{d}_m$, related to the Fourier image of the original filtering function, take values,

$$\begin{aligned} \mathrm{d}_m &= [\mathrm{d}_{-3}, \mathrm{d}_{-2}, \mathrm{d}_{-1}, \mathrm{d}_0, \mathrm{d}_1, \mathrm{d}_2, \mathrm{d}_3] \\ &= [-1, 6, -15, 20, -15, 6, -1]/64. \end{aligned} \tag{88}$$

The application of the filter at a given time step can be considered as changing a variable by,

$$\frac{\partial u}{\partial t} = \frac{u(x) - u_{\text{filt}}(x)}{\Delta t} = \sum_{m=-3}^{3} \mathrm{d}_m u(x + m\Delta x)/\Delta t. \tag{89}$$

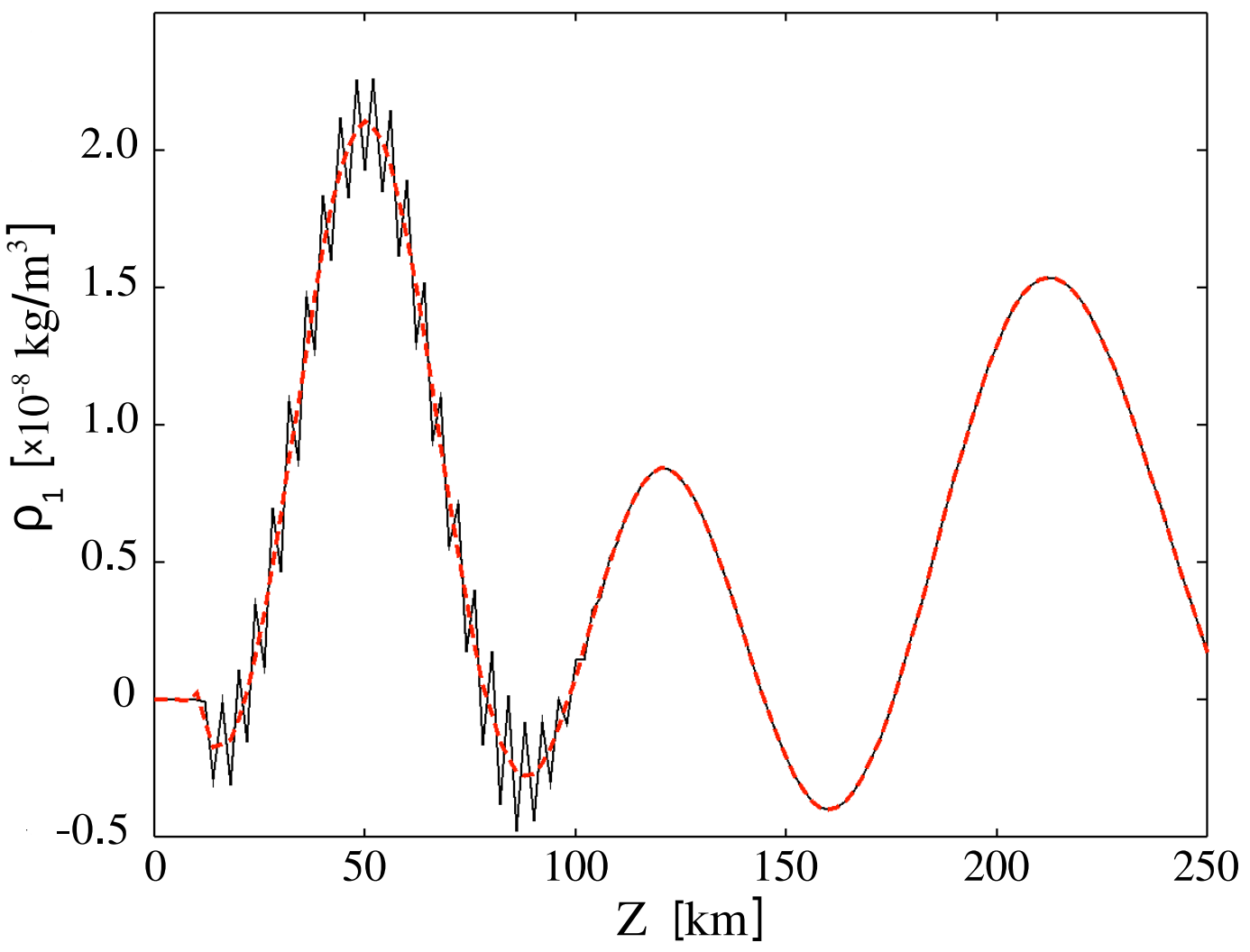

**Figure 3.** The second-order density perturbation as a function of height, in a 1D experiment of a monochromatic Alfvén wave propagation from the bottom photosphere upwards in a stratified solar atmosphere permeated by a constant magnetic field. Solid black line: no filtering is applied. Dashed red line: filtering applied.

Therefore, filtering operation can be viewed as an additional type of the 6th order diffusion, (see the Appendix in Popescu Braileanu, Lukin, and Khomenko, 2023),

$$\frac{\partial u}{\partial t} = \nu_6^F \nabla^6 u, \tag{90}$$

where $\nu_6^F$ is the diffusion coefficient. In the finite difference representation the 6th derivative of $u$ on the seven-point stencil is approximated with,

$$\nabla^6 u = \frac{\partial^6 u}{\partial x^6} = \frac{1}{\Delta x^6} \sum_{m=-3}^{3} \mathrm{c}_m u(x + m\Delta x), \tag{91}$$

with $\mathrm{c}_m = 64\mathrm{d}_m$. Then the diffusion coefficient $\nu_6^F$ introduced by the filter can be evaluated as,

$$\nu_6^F = \frac{\Delta x^6}{64\Delta t}, \tag{92}$$

where $\Delta x$ is the grid spacing along $x$.

Figure 3 demonstrates the main effect of filtering. It shows the second-order density perturbations appearing as a consequence of the non-linear coupling in the experiment where an Alfvén wave was excited at the bottom photospheric boundary of a 1D simulation domain (see Appendix 5 in Felipe, Khomenko, and Collados, 2010, for the details of this test). In the absence of the filtering (black curve), there is a high-frequency point-to-point noise visible over the main oscillation. The noise amplitude is affected by the constant artificial diffusion (see the Alfven speed contribution in Eq. 83), which has the lowest value at the bottom of the domain and grows with height due to exponential profile of the background density. The filtering removes numerical noise while preserving the main oscillation and its amplitude (red curve). On the other hand filtering should not be applied too often, as it introduces small inconsistency to the governing equations and may also overshoot original solution in case of large gradients like shocks. As a rule, applying filtering every 20–50 iterations produces stable simulations maintaining physical gradients and diminishing noise in the solution. Finally, it is interesting to notice that the filtering fits perfectly the concept of split variables. According to Hesthaven (1998), a high frequency noise filtering can improve the long-time stability of the PML, see Sect. 3.7. In Mancha3D this approach is shown to work successfully for simulations of MHD waves.

### 3.6. Divergence of magnetic field

In Mancha3D no specific treatment is applied to control the divergence of the magnetic field. The split variable strategy greatly facilitates maintaining the divergence-free condition. In the simulations with non-zero equilibrium background, the $\nabla \cdot \mathbf{B} = 0$ is analytically fulfilled in the initial state and it remains so through the whole simulation since the initial state is not evolved. Since the code operates in perturbations, it allows keeping $\nabla \cdot \mathbf{B} = 0$ condition to the zero order. We also find that our centered numerical scheme allows keeping $\nabla \cdot \mathbf{B} = 0$ to a good degree of precision in the simulations with full variables (zero background).

In order to check the behaviour of $\nabla \cdot \mathbf{B} = 0$ we performed a 3D simulation of the Rayleigh-Taylor instability (RTI), where a dense plasma lies on top of a less-dense plasma with typical solar atmosphere parameters for the density and temperature. We use a relatively small setup of $120 \times 120 \times 320$ grid points in x-y-z directions with $h = \Delta x = \Delta y = \Delta z = 10$ km. Using the ideal gas equation, the stable configuration in a gravitational field corresponds to the adiabatic profiles of density, pressure and temperature,

$$\rho = \rho_0 \left( 1 - \frac{\gamma - 1}{\gamma} \frac{\rho_0 g}{p_0} z \right)^{1/(\gamma-1)}, \quad (93)$$

$$p = p_0 \left( 1 - \frac{\gamma - 1}{\gamma} \frac{\rho_0 g}{p_0} z \right)^{\gamma/(\gamma-1)}, \quad (94)$$

$$T = T_0 \left( 1 - \frac{\gamma - 1}{\gamma} \frac{\rho_0 g}{p_0} z \right), \quad (95)$$

where $\mathrm{g} = 274\,\mathrm{m\,s^{-1}}$ is the gravity at the Sun surface, $\gamma = 5/3$ is the adiabatic factor, $\rho_0 = (\rho_{\mathrm{top}} + \rho_{\mathrm{bot}})/2$, $T_0 = 10^4\,\mathrm{K}$, and $p_0 = (R/\mu_{\mathrm{g}})\rho_0 T_0$ are the reference density, temperature and pressure, $\rho_{\mathrm{bot}} = 10^{-8}\,\mathrm{kg\,m^{-3}}$, $\rho_{\mathrm{top}} = 10^{-7}\,\mathrm{kg\,m^{-3}}$, $\mu_{\mathrm{g}} = 10^{-3}\,\mathrm{kg\,mol^{-1}}$ is the molar mass. The RTI is triggered by a half-cosine perturbation of vertical velocity set at the initial interface between the heavy and light plasmas. It leads to a single uprising bubble in the middle of the domain and following spikes at its sides, as shown in Figure 4, top left panel. Initially there is no magnetic field in the setup. The field is generated by the Biermann battery mechanism, as the instability evolves (Khomenko *et al.*, 2017; Martínez-Gómez *et al.*, 2021).

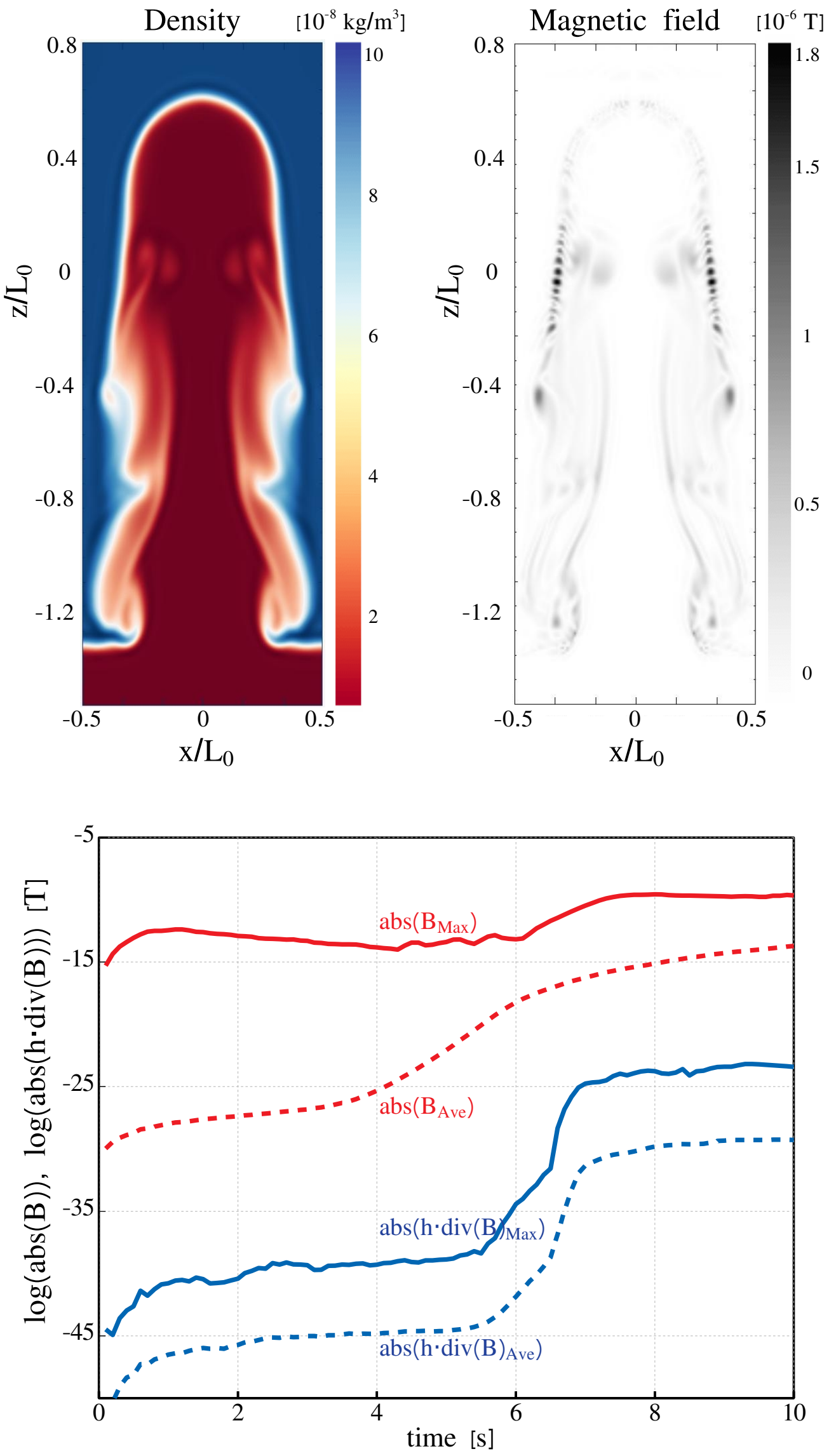

**Figure 4.** Top panel: density (left) and magnetic field (right) distributions in a simulations of Rayleigh-Taylor instability at time=5.5 s. Bottom panel: time-evolution of the maximum and volume averaged values of the magnetic field and its absolute divergence multiplied by the grid step, *h*, in logarithmic scale.

Figure 4 shows the distributions of the density and the generated magnetic field in the nonlinear stage of the instability. At the bottom panel, we plot the time evolution of the maximum and volume averaged values of the magnetic field module and its divergence in a logarithmic scale; the magnetic field divergence is taken as its absolute value multiplied by the grid spacing, so that it has the units of magnetic field strength. First of all, this plot does show the existence of non-zero magnetic field divergence. At the early stage of the instability, $\nabla \cdot \mathbf{B}$ grows together with the growth of the magnetic field. Around the time of 6 seconds, the bubble reaches the top boundary. By that time the flow becomes much more turbulent leading to further generation of the magnetic field by both Biermann battery and local dynamo effects. The sharp increase in the $\nabla \cdot \mathbf{B}$ terms is determined by the non-periodic boundary condition at the top of the domain, which is in this case modelled as a simple insulator and produces most of the magnetic field divergence. It is important to notice, that in the highly turbulent stage ($t > 7$ seconds) the growth of the divergence saturates both in its maximum and its average values. Furthermore, the absolute values of $h\nabla \cdot \mathbf{B}$ is about 10 orders of magnitude smaller than the corresponding value of the generated magnetic field, which clearly indicates that its effect on the flow evolution is negligible.

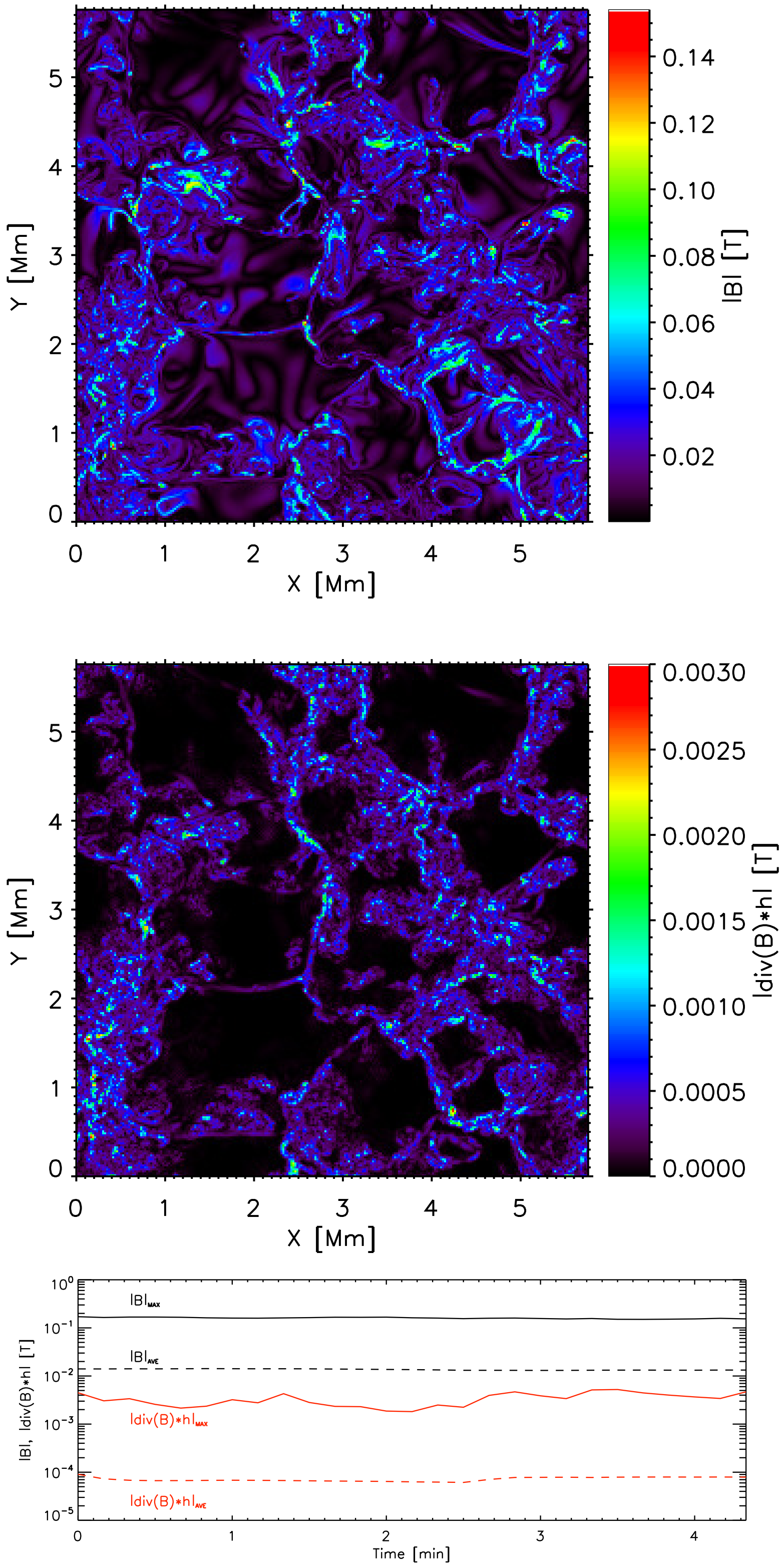

**Figure 5.** Magnetic field modulus (top), and its absolute divergence (middle) distributions over the horizontal surface at photospheric base, in realistic simulations of small-scale solar dynamo, as those reported in Khomenko *et al.* (2018) but with a horizontal/vertical resolution of 5/3.5 km, respectively. Bottom panel: time-evolution of the maximum and volume averaged values of the magnetic field and its absolute divergence multiplied by the grid step, *h*, in logarithmic scale.

We also computed the magnetic field divergence in a time series of realistic simulations of small-scale solar dynamo, similar to those reported in Khomenko *et al.* (2018), but done at a different spatial resolution of 5/3.5 km horizontally/vertically. In this simulation the magnetic field was initially seeded through the Biermann battery term in in the induction equation and it was then amplified by the action of dynamo in the near surface layers (Khomenko *et al.*, 2017). An example of this computation is given in Figure 5. The upper panel reveals a typical network-like pattern of small-scale magnetic structures coinciding with intergranular lanes at the solar surface, with a mean strength slightly above 0.012 T. Similarly to the RTI case, the middle panel shows a presence of a non-zero divergence. The locations of highest divergence correlate with the locations of the highest magnetic field. The divergence values in this case are not as small as for the RTI simulation, but nevertheless they keep at the value of $\approx$2 orders of magnitude smaller than the magnetic field generated by the small scale dynamo. The value of the divergence is constant in time in the stationary phase, see the bottom panel. Given the complexity of this realistic simulation, compared to the case of the RTI, we consider these values acceptable.

### 3.7. Split variables and PML

Splitting variables into equilibrium and perturbation parts stems from the very origin of the MANCHA3D code to study wave propagation in the solar atmosphere (Khomenko and Collados, 2006; Felipe, Khomenko, and Collados, 2010). It allows preserving potentially large contrast between oscillating and background quantities with a good numerical accuracy. In fact, any setup can be run with either split or full variables. In the latter case the equilibrium is set to zero and everything is computed as the perturbation part. The structure of the equations solved by MANCHA3D is specifically designed to use the split variables together with the PML boundary conditions (Berenger, 1994; Hu, 1996; Parchevsky and Kosovichev, 2007). These PML boundaries are very effective for wave simulations to prevent spurious wave reflections, absorbing perturbations within about 10–20 grid points. The PML can be used at all the boundaries of the computational domain.

Playing a role of non-reflecting boundaries the PML should not be considered as a type of boundary conditions as it usually takes more grid points and it requires solving the modified governing equations in the PML. Following the PML strategy, the MHD equations in MANCHA3D are reshaped in order to add a term that damps the perturbations that reach the boundary, separately for each direction. In a 3D geometry, the schematic representation of the system of equations, Eq. 50, is expanded as follows,

$$\frac{\partial \mathbf{u}}{\partial t} + \frac{\partial \mathbf{H}(\mathbf{u})}{\partial x} + \frac{\partial \mathbf{G}(\mathbf{u})}{\partial y} + \frac{\partial \mathbf{K}(\mathbf{u})}{\partial z} = \mathbf{S}(\mathbf{u}), \tag{96}$$

where $\mathbf{u} \equiv [\rho_1, \rho\mathbf{v}, e_1, \mathbf{B}_1]$ is the vector that contains the conserved variables; the vectors $\mathbf{H}$, $\mathbf{G}$ and $\mathbf{K}$ are the fluxes written separately for each direction, and $\mathbf{S}$ represents the source terms. The conserved variables $\mathbf{u}$ are split into three components in such a way that $\mathbf{u} = \mathbf{u}_1 + \mathbf{u}_2 + \mathbf{u}_3$ and also $\mathbf{S}(\mathbf{u}) = \mathbf{S}_1(\mathbf{u}) + \mathbf{S}_2(\mathbf{u}) + \mathbf{S}_3(\mathbf{u})$ and the system of MHD equations is split into a set of three coupled, one dimensional equations:

$$\frac{\partial \mathbf{u}_1}{\partial t} + \frac{\partial \mathbf{H}(\mathbf{u})}{\partial x} + \sigma_x(x)\mathbf{u}_1 = \mathbf{S}_1(\mathbf{u}), \tag{97}$$

$$\frac{\partial \mathbf{u}_2}{\partial t} + \frac{\partial \mathbf{G}(\mathbf{u})}{\partial y} + \sigma_y(y)\mathbf{u}_2 = \mathbf{S}_2(\mathbf{u}), \tag{98}$$

$$\frac{\partial \mathbf{u}_3}{\partial t} + \frac{\partial \mathbf{K}(\mathbf{u})}{\partial z} + \sigma_z(z)\mathbf{u}_3 = \mathbf{S}_3(\mathbf{u}), \tag{99}$$

where $\sigma_{x,y,z}$ are damping coefficients along each direction. Notice that in the case of solving 2D equations, the splitting is done into 2 components, but the same philosophy applies otherwise. In the 1D case, the PML layer simply converts into a usual absorbing sponge layer. The damping coefficients are non-zero only in the PML part of the domain, and are zero in the physical domain.

Theoretically, a PML with constant damping $\sigma_{x,y,z}$ does not produce spurious reflections for the incident plane waves for any angle of incidence and at any frequency. However, numerically, reflections may appear when $\sigma_{x,y,z}$ have a steep gradient (Berenger, 1994). To solve this problem MANCHA3D includes smooth variations in the absorption coefficients from small values at the interface between the PML medium and the physical domain to large values at the outer boundary,

$$\sigma_x = a_x \frac{c_{S0} + v_{A0}}{\Delta x} \left( \frac{x - x_{\rm PML}}{x_{\rm PML}} \right)^2, \tag{100}$$

$$\sigma_y = a_y \frac{c_{S0} + v_{A0}}{\Delta y} \left( \frac{y - y_{\rm PML}}{y_{\rm PML}} \right)^2, \tag{101}$$

$$\sigma_z = a_z \frac{c_{S0} + v_{A0}}{\Delta z} \left( \frac{z - z_{\rm PML}}{z_{\rm PML}} \right)^2, \tag{102}$$

where $a_x$, $a_y$ and $a_z$ are constants controlling the damping amplitude, and $x_{\rm PML}$, $y_{\rm PML}$ and $z_{\rm PML}$ are the thickness of the PML domain in each spatial direction. In a typical calculation, MANCHA3D needs a PML of 10 – 20 grid points. The coefficients $a_x$, $a_y$ and $a_z$ depend on each particular simulation, and vary between 0 and 1. For low frequency waves, corresponding to longer wavelengths, MANCHA3D requires wider and weaker PMLs, compared to the high-frequency waves. When the vertical wavelength of the wave becomes comparable or larger to the size of the PML, it becomes harder to absorb and eventually may produce reflections (such situation is typical in coronal conditions where Alfvén waves are present). Too wide or too strong PMLs can become numerically unstable. The PML formulation can also become unstable in long simulation runs because of the accumulation of the high-frequency noise coming from waves propagating tangentially to the boundary.

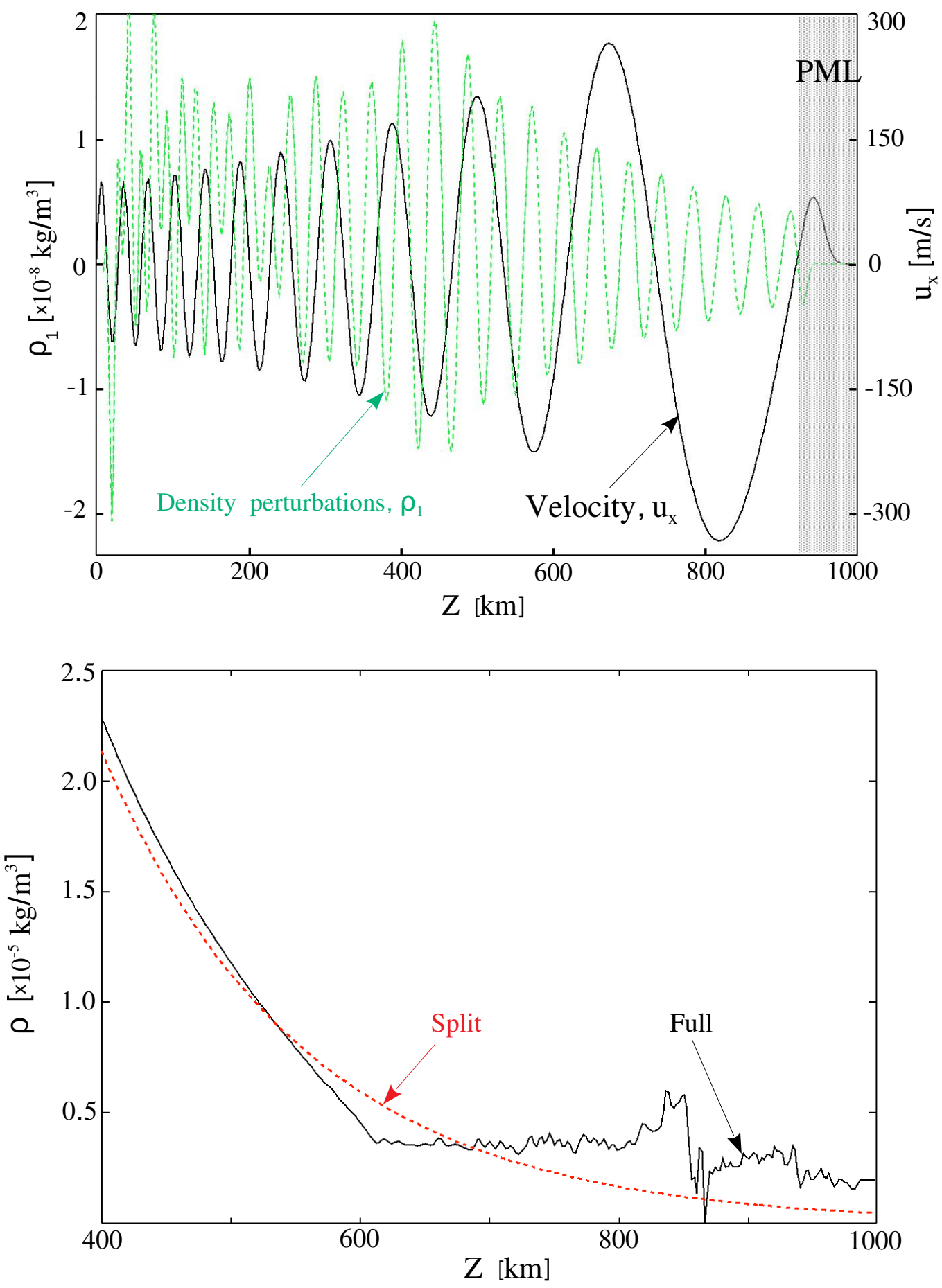

**Figure 6.** Top: profiles of the density perturbations and horizontal velocity as a function of height in a simulation of a linearly polarized monochromatic Alfvén wave propagation in a stratified solar-like atmosphere, permeated by a vertical constant magnetic field of a strength $B_Z$ = 500 G. Bottom: total density profile plotted for the upper part of the domain for the same setup computed using the split (red dotted line) and full (solid black line) variables.

We illustrate the advantage of the split variables and the PML layer by using the simulations of linearly polarized Alfvén wave propagation in a stratified solar atmosphere permeated by a constant vertical magnetic field, $B_z$ = 500 G; the wave is triggered by a velocity oscillation at the bottom of the domain with period of 10 s and amplitude of 100 m s$^{-1}$. The top panel in Figure 6 shows a snapshot of the density and velocity, as a function of height, together with the location of the PML at the top boundary of the computational domain (grey shaded area). Perturbations in density appear as a second-order effect due to the nonlinear evolution of the MHD equations. The background density varies from $2.7 \cdot 10^{-4}$ kg m$^{-3}$ at the bottom to $4.7 \cdot 10^{-7}$ kg m$^{-3}$ at the top boundary, while the amplitude of the density perturbations do not exceed $2 \cdot 10^{-8}$ kg m$^{-3}$, i.e., $10 - 10^4$ times smaller than the equilibrium values. It is easily seen that the PML (which in 1D case coincides with a sponge layer) effectively suppresses all the perturbation at the top boundary, allowing for a stable long simulation.

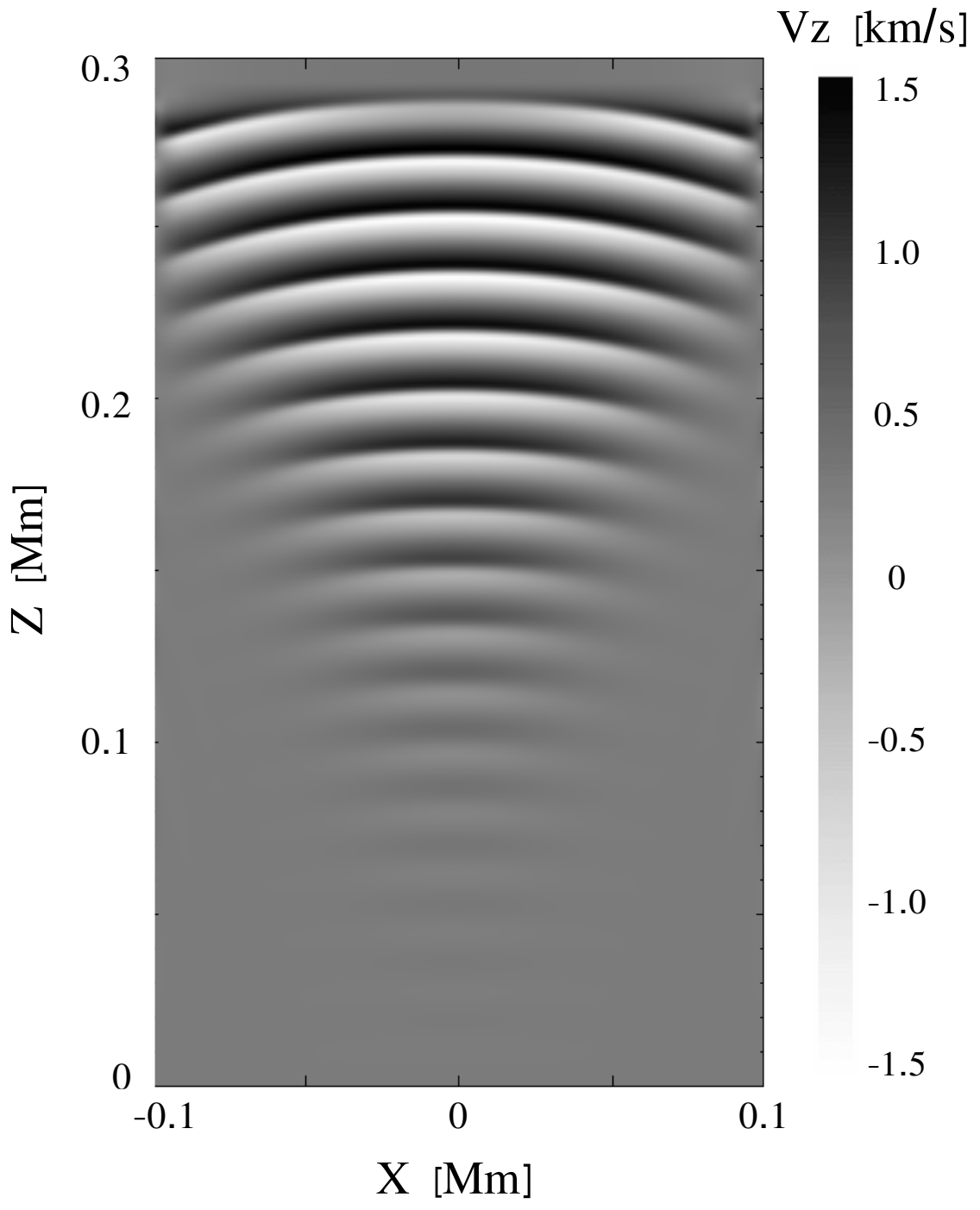

**Figure 7.** 2D snapshot of the vertical velocity in the simulation of acoustic wave propagating through the solar-like atmosphere with a PML boundary conditions on the top and on the sides.

The same simulation can be set up with the full variables and a simple outflow boundary condition at the top, as the PML boundary condition cannot be applied in this case. However, such a configuration crashes when the wave approaches the top part of the domain with a relatively small density. The total density profile for this case is shown by a black solid line at the bottom panel of Figure 6, together with the same quantity for the simulation with the split variables (red dotted line). The unstable evolution in this case is attributed to the fact that the numerical treatment leads to small growing deviations from the equilibrium even if the initial MHS profile is set analytically. Consequently, in the full variable case, the total density inevitably evolves departing from its equilibrium state. Furthermore, possible reflections from the top boundary may also affect the flow, so that at some point the density may become negative and the simulation crashes.

The action of the PML layer in a 2D configuration is demonstrated by an example of a monochromatic acoustic wave propagation through a solar-like stratified atmosphere. The equilibrium profiles are the same as in the previous Alfvén test, but without magnetic field. The wave is triggered in the middle of the bottom boundary with 20 s period and 10 m $s^{-1}$ initial amplitude; the top boundary condition has 25 grid points PML and the sides have 15 grid points PML. Figure 7 shows the 2D distribution of the vertical velocity at time moment t = 450 s, when the wave propagates in a quasi-steady way. It clearly demonstrates that all the perturbations at the domain boundaries are successfully damped by the PML.

**Table 2.** PizDaint node characteristics. L1 and L2 caches belong to each processor while L3 cache is shared within a socket (6 CPUs) and RAM is shared within a node

| Number of CPU | Number of sockets | L1 cache | L2 cache | L3 cache | RAM | CPU speed |
|---|---|---|---|---|---|---|
| 12 | 2 | 64 kB | 256 kB | 30 MB | 192 GB | 2.3 GHz |

## 4. Parallel efficiency

Parallel performance is an important issue of every modern computer code, however it is rarely discussed in regular scientific papers. MANCHA3D is fully parallelized with the MPI standards for distributed memory machines (Hager and Wellein, 2011). The computation domain can be decomposed in all three directions. The output files of the HDF5 format (The HDF Group, 2000-2010) are read and written in parallel by all the processors.

We explore the parallel efficiency of the code by using a simulation of the Kelvin-Helmholtz instability with two counter-directed flows. This setup has the advantage to be easily scalable in all three dimensions. For the sake of saving computational resources, all the tests are performed without non-ideal effects. As a rule, the inclusion of the ambipolar diffusion or the Hall term can increase the overall computational time, however the additional treatment of those terms is not expected to noticeably affect the parallel efficiency.

It should be emphasized that the obtained results are specific for a particular machine, with its unique configuration of the CPUs, cache memory size, bandwidth, file system, etc. All the tests have been performed on the PizDaint supercomputer of the Swiss National Supercomputer Center; its details are summarized in Table 2. Nevertheless, they provide a general understanding of the code performance and expectation of its behavior on other machines. Below we discuss in detail the strong and weak scaling of the code including the effect of saving snapshots.

### 4.1. Strong scaling

Strong scaling implies a fixed size problem run on a single processor and compared to running on many processors (Gustafson, Montry, and Benner, 1988). The main purpose of the strong scaling tests is twofold: to demonstrate parallel efficiency and to determine the optimal size of the decomposition subdomain for further simulations. By default the code speedup is usually estimated as:

$$\text{Speedup} = \frac{\mathrm{t_1}}{\mathrm{t_N}}, \tag{103}$$

where $t_1$ is the wall-clock time for running the code on a single CPU and $t_\mathrm{N}$ is the time for running the same setup on N processors. In ideal case, for instance, doubling the number of processors decreases the code execution time by 50 percent. There are several technical issues to keep in mind while performing strong scaling tests:

- Due to the limited memory per CPU, it is impossible to run a relatively big size job on a single CPU.
- The possible range for an efficient number of processors becomes limited, as the decomposition subdomain size should be considerably larger than the number of cells to exchange.
- The architecture of a modern supercomputers implies a large number of nodes, each of them with many CPUs, where the actual computation takes place; the number of CPUs per node varies from a few to several dozen. For a proper scaling it is important to avoid the comparison of the code performance within the node with the one across the nodes. For this reason, scaling tests are performed with respect to a single node (or even several nodes for big setups) and not a single CPU.
- The strong scaling efficiency may exhibit non-smooth behavior due to varying relation between the amount of memory required for a subdomain and the CPU cache memory. As a result, it becomes very much machine dependent. Furthermore, different machines have different sets of compilers, different hardware connections which inevitably affect running the code as a serial or as a parallel version.
- As a rule no snapshots are saved in the scaling tests in order to exclude parallel input/output which is a separate issue.

Keeping in mind the above reasoning we performed several sets of parallel runs with different 2D setups. In all the cases, the simulation runs for 10000 iterations with a different number of CPUs. No snapshots are saved in these tests as we are interested in the parallel efficiency of the code and want to exclude other hardware effects like I/O speed. Figure 8 shows the results for the strong scaling test, for different sizes of the computational domain ($2880 \times 1800, 4320 \times 4800$ and $1152 \times 900$ points, from top to down). The left vertical axis shows the subdomain size computed by each CPU and the right axis shows the actual parallel speedup; the x-axis stands for the number of CPUs.

We start our analysis from the top panel of Figure 8 with the setup of medium domain size. The speedup is scaled with respect to 12 CPUs which corresponds to a single full cluster node. Here the strong scaling reveals several interesting aspects in its behaviour. The first striking feature is that for a relatively small number of CPUs ($< 2000$) the code overperforms the ideal scaling exhibiting superlinear speedup (Gusev and Ristov, 2014). For example, when the number of CPUs doubles from 48 to 96 the wall-clock time drops 2.6 times. It should be noted that the overperformance is related to the slope of the curves (the red and the blue ones) and to the fact that one curve is above the other one. Starting from around 1000 CPUs due to increase in the number of communications between CPUs the parallel efficiency is degrading and its slope becomes less than 1. At about 2400 CPUs there is an inflection point followed by a counterintuitive growth in the parallel speedup, almost reaching the ideal slope. Finally with the further increase of the CPUs number the communication cost grows significantly and speedup becomes negative.

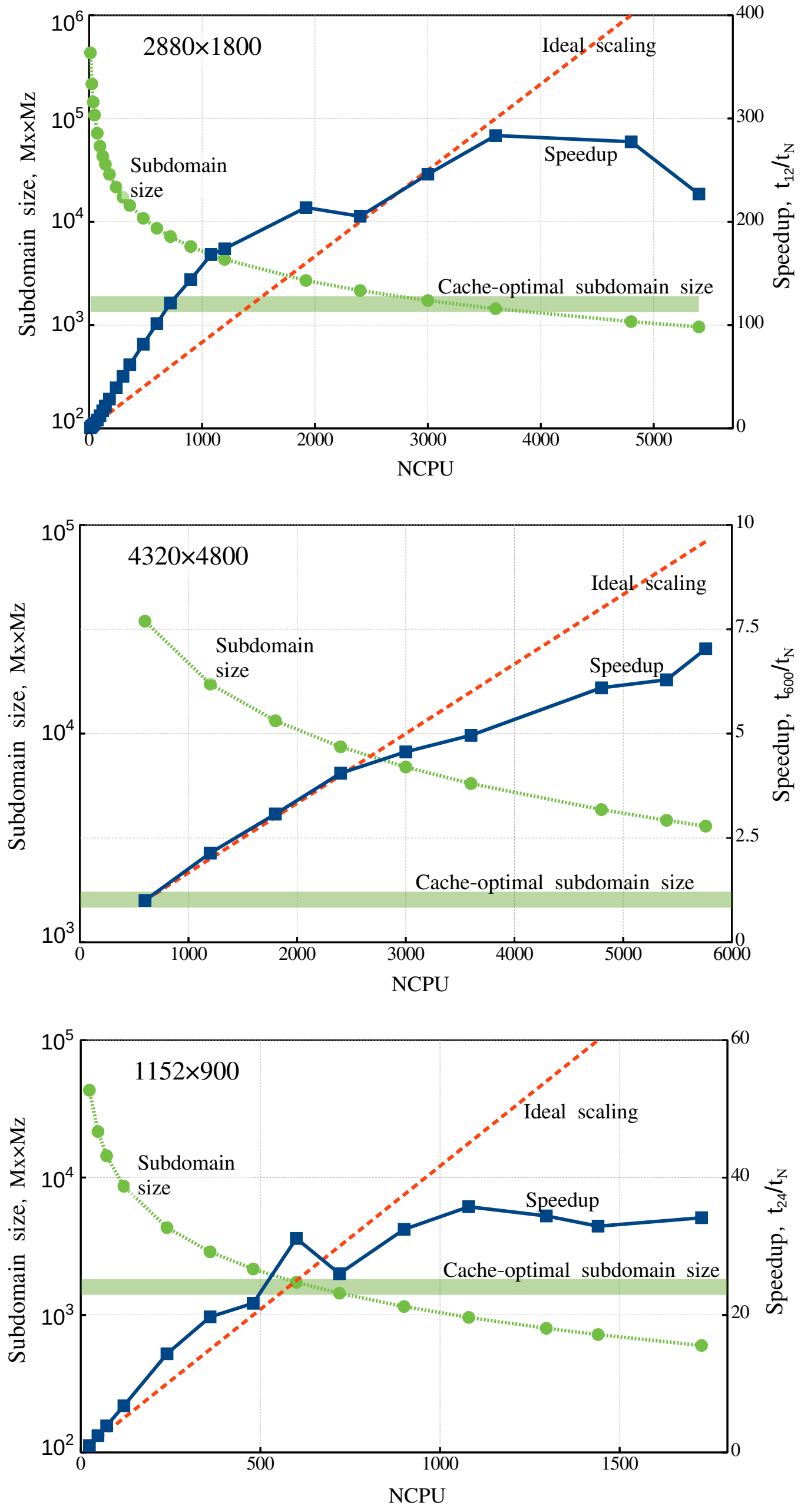

**Figure 8.** Strong scaling for 2D setup of different sizes: medium (2880×1800) - upper panel, large (4320×4800) - middle panel and relatively small (1152×900) - low panel. Solid blue squares lines show the numerical results, dashed red lines stand for the ideal scaling, green dotted lines with circles represent the subdomain size, and the shaded areas depict an optimal subdomain size with respect to the CPU cache memory.

In order to get a better insight the superlinear speedup and to explain the inflection point, it is necessary to know the internal architecture of the computer node. Each node contains several processors, each processor has its own L1 and L2 cache memories, and the L3 cache memory is shared among several processors within one socket and finally the RAM which is common for all the processors of the node. The L1 and L2 caches are fast but small memories, usually in total less than 300-500 Kb, while the L3 cache is much larger, 20-30 Mb. When a processor needs specific data, it searches for it in the L1 cache, then in the L2 cache, in the L3 cache and finally in the RAM. The access time increases by order of magnitude while addressing the L3 cache as compared to the L1 cache and addressing the RAM takes much longer (Stengel *et al.*, 2015). When several CPUs work in parallel, they are competing for getting the memory access in the L3 cache within each node. Consequently, the larger is the subdomain handled by each processor, the harder is to get an access to the data for all the processors of the compute node. As the subdomain size decreases, more and more processors are able to access the L3 cache memory simultaneously, which, in turn, decreases the overall wall-clock time. In Figure 8 we depict the subdomain size treated by each CPU (the left axis and the green circle dotted curve) to show its correlation with the code parallel efficiency. Initially (left side of the plot with a very few nodes), each CPU needs more memory than available at the L3 cache (which is shared by several CPUs), and it frequently accesses "slow" RAM. In the worst case the CPUs within one socket work in a serial way rather than in parallel. Strictly speaking, under such conditions the code underperforms within each node but overperforms in overall. Reducing the subdomain size leads to the situation when more and more CPUs can access the fast L3 cache in parallel, and explains the initial code overperformance.

The noticeable speedup growth after the inflection point at about 2400 CPUs means that the computational subdomain treated by a node becomes small enough so that it fits the L2+L3 cache memories, and at the same time it is still large enough to exceed the exchanges between CPUs. In particular, on PizDaint machine, each processor has about 2.5 Mb of the L2+L3 cache memories, which corresponds to the subdomain size $50\times50$ grid points treated by each CPU (the green area in Figure 8). So when the subdomain size is $50\times50$ or below, all the CPUs of a node are able to treat their data simultaneously. For larger number of CPUs, the communication overhead decreases the overall parallel efficiency and the strong scaling shows the saturation and even degrading behaviour.

All the above clearly indicates that the code is strongly memory bound (Stengel *et al.*, 2015). It means that the CPU speed is not the limiting factor for effective parallel computation. It also reduces the impact of MPI parallelization and exchange between subdomains. For Mancha3D, high memory demands stem from the usage of a large number of additional variables needed for the split variables and the PML treatment. Nevertheless, such a behaviour should be inherent to most MHD codes regardless of the numerical algorithms used.

We have performed two more sets of runs to clarify the reasons of parallel efficiency degradation. The middle panel of Figure 8 shows the result for a larger 2D domain, 4320 × 4800; such a big domain does not fit a single node and the speedup is scaled with the computing time for 50 nodes (600 CPUs). One can see that even with 5760 CPUs the subdomain size treated by each processor is larger than the optimal one. Due to communications the parallel efficiency drops just to 70% which is still a very good result for several thousand CPUs run. If we keep increasing the number of processors the speedup will eventually become negative.

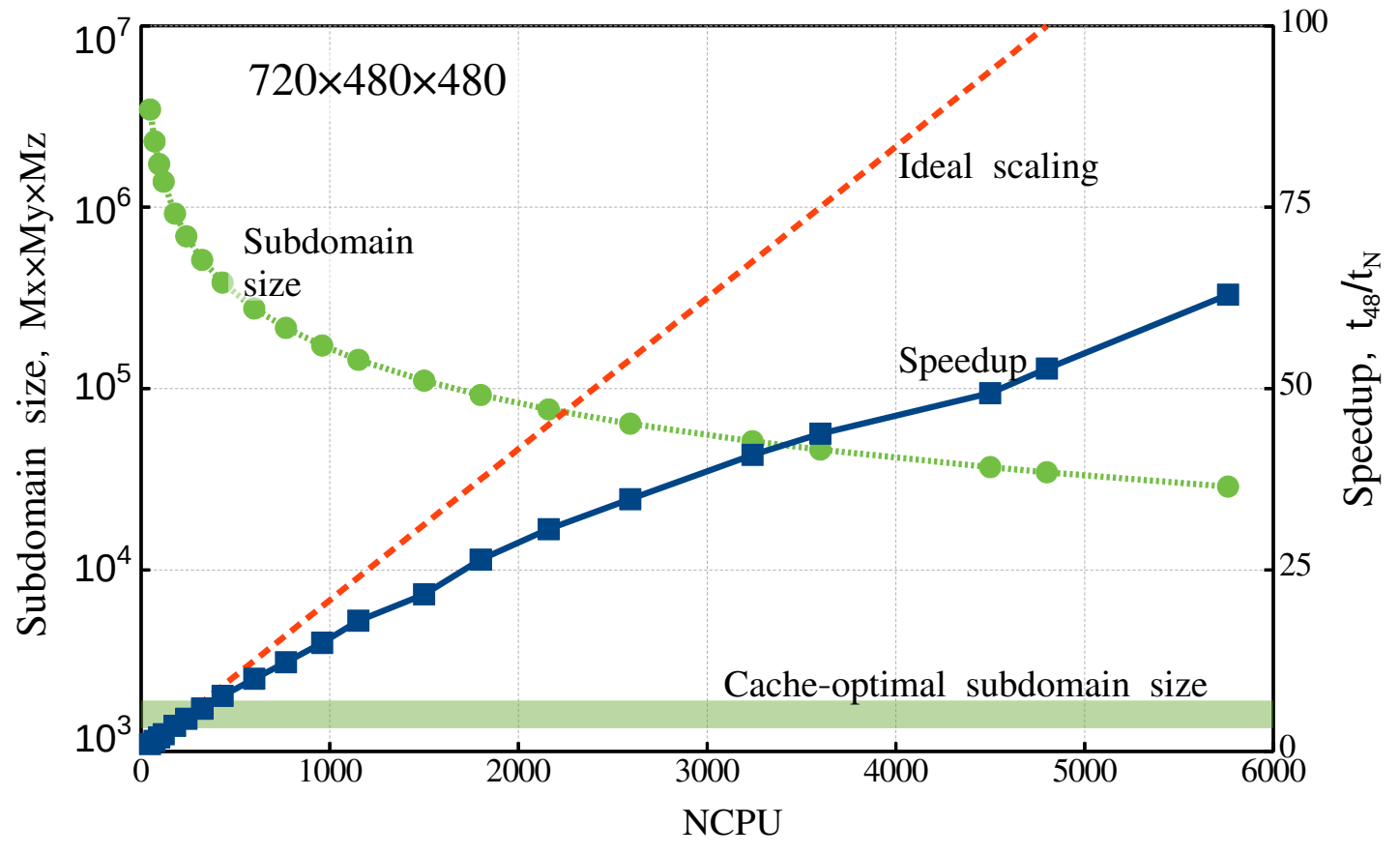

**Figure 9.** Strong scaling for a 3D setup. The format of the figure is the same as Figure 8.

The bottom panel of the Figure 8 shows the opposite extreme when the computational setup is relatively small, 1152 × 900. The parallel speedup overperforms for a small number of CPUs and reaches its peak around 600 CPUs, when the subdomain size fits the node cache memory. With further increase of the CPUs the subdomain becomes smaller increasing the relative amount of exchanged data between CPUs. It leads to a strong degradation of the parallel speedup already at 1000 CPUs.

Finally in Figure 9 we show the strong scaling for a 3D setup in the same format. The cache-optimal size remains the same and, inevitably for 3D, stays well below any possible subdomain size. In the 3D setup the number of communication grows as there is one extra dimension for the data exchange, and the speedup degrades faster than the one of the big 2D setup. The overall speedup drops by 30% at 2000 CPUs and exceeds 50% at 5000-6000 CPUs, which is an acceptable code performance[4]. It is important to notice that the smallest subdomain size corresponds to $30^3$; decomposing to smaller sizes is possible but it will be less efficient due to the increase in communication overheads.

### 4.2. Weak scaling

Another typical situation in scientific computing is to increase the computational domain while keeping the same resolution. The code performance for a fixed domain size per processor but increasing the overall size of the problem and the number of processors is referred to as the weak scaling. Unlike the strong scaling, the weak scaling shows how the code parallel performance degrades when the number of processors increases, regardless the subdomain size treated by individual processor. Ideally the execution time should not change with the increase of the number of the processors, but due to communications between processors and other cache memory issues there will be inevitably some increase in overall wall-clock time. The weak scaling efficiency can be evaluated in the same way as the speedup given by Eq. 103.

[4]For example, for PizDaint applications it is required to show parallel efficiency of at least 50%.

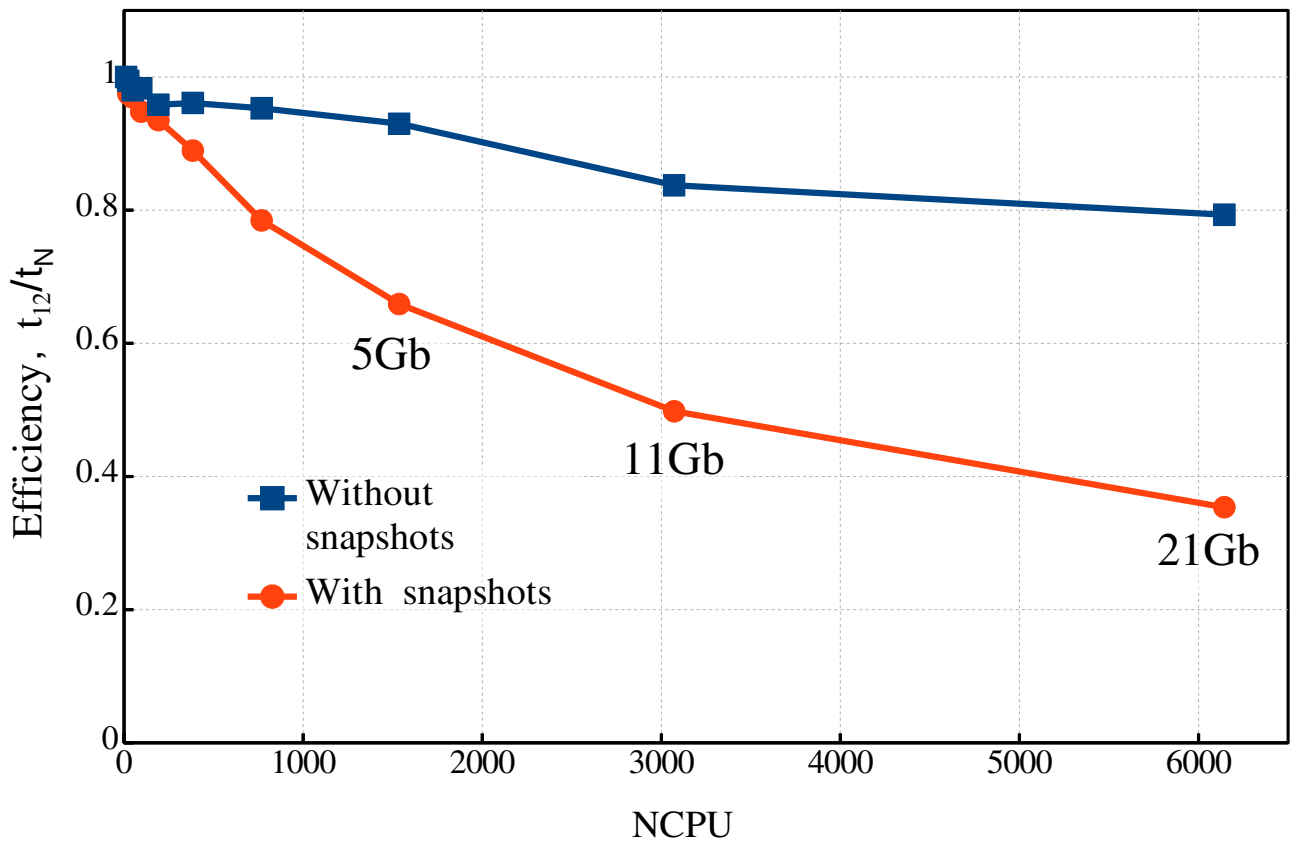

**Figure 10.** Weak scaling efficiency for a 3D setup with a subdomain size of $40 \times 40 \times 40$ points. The two curves correspond to running the jobs with and without saving snapshots, the numbers stand for the snapshot size.

Figure 10 presents a weak scaling for a 3D setup with a moderate subdomain size. The blue squares curve shows an impressive weak scaling, degrading only by 20% at more than 6000 processors. The weak scaling efficiency demonstrates monotonous behaviour as compared to the strong scaling, it does not depend on the subdomain size as each node has the same load regardless the total CPU number.

Saving snapshots is an inevitable part of every simulation. We have performed additional runs to get a better insight into how the snapshot writing affects the code's behaviour. In general, the saving delays depend on the snapshot size, the saving frequency, file system, etc. The red circles line depicts the case when the snapshots are saved every 25 iterations, leading to a significant increase of the overall wall-clock time. Such a behaviour should be inherent to weak scaling tests, since an increase in the number of processors leads to a corresponding increase in the total domain size of the problem. As expected, when the snapshot size is relatively small ($< 1$ Gb), the delays due to its saving are negligible, meaning that it takes much less time than the computations. In MANCHA3D the snapshot size of 1 Gb corresponds, for example, to a large 2D setup, like $4000 \times 4000$ points, or to a moderate 3D domain of $400 \times 400 \times 100$ points. For a bigger computational domain, for example, $800 \times 400 \times 500$, the snapshot size exceeds 10 Gb and it takes $\sim 20$ seconds to save it on the PizDaint machine. Hence, if it is necessary to save such snapshots very frequently, the writing time can become the dominant one in the overall execution time. In particular, for the dimensions mentioned above, saving snapshots every 25 iterations doubles the overall time of the simulation, as compared to the run without snapshots.

Thus we conclude that the MANCHA3D code is effectively scalable up till several thousands of the CPUs. For 2D simulations there is an optimal subdomain size allowing significant gain in the parallel performance. For 3D cases it is more efficient to run the code with larger subdomains, avoiding sizes smaller than $25 - 30$ grid points in each direction. It should also be kept in mind that frequent saving of big snapshots increases the overall running clock time, regardless the number of processors used in the simulation.

## 5. Conclusion and perspectives

In this paper we describe in detail the equations and numerical methods used in the MANCHA3D code. Special attention has been paid to specific features of the code, like split variables, the PML boundary condition and the enhanced stability due to the filtering technique used. Several test simulations have been used to demonstrate the performance of these specific features. We note that the overall performance of the code's numerical scheme has been reported in Felipe, Khomenko, and Collados (2010), and, therefore we do not repeat those tests in the current paper. Neither we discuss scientific simulations, since all of the results have been widely reported, as discussed in the Introduction. These simulations range from those of helioseismic wave propagation below sunspots, non-ideal effects due to neutrals, large and small-scale prominence dynamics, or waves in coronal structures. As the current paper shows, the code shares several features with other widely used numerical codes, but it also has its unique features, and it has been specially fine-tuned for simulations of wave propagation in static magnetic structures and for realistic simulations of magneto-convection.

The parallel efficiency of the code has been thoroughly studied in this paper for the first time. It has revealed a non-trivial strong scaling behaviour for 2D setups, which has been properly explained from the hardware viewpoint. The weak scaling exhibits almost perfect properties, although it is affected by the snapshot saving frequency, due to significant large file size, especially in 3D simulations. Both scalings clearly show that the code is memory bound, which should be a common property of MHD codes. As a consequence of being memory bound, the strong scaling has a different behavior for 2D and 3D setups. For the former (smaller) case there is an optimal subdomain size of $\sim 2000$ grid points which allows fitting the data to the CPU memory with reasonable exchange load between subdomains. On the contrary, for the 3D setup one should keep the subdomains larger than 30x30x30 grid points to avoid high exchange overheads. In this case the code can effectively run on several thousand CPUs.

The code is still expanding with new features. The thermal conduction module, described in Sect. 2.1, is one of the newest modules that has been added allowing modeling the dynamics of the hot solar corona. The static non-uniform grid in the vertical direction, described in Sect. 3.2.2, is another recent implementation that will also contribute to the studies of the solar corona as well as other simplified setups, where there is a need for a larger domain. The long term upgrades involve including a more complex model for chromospheric radiative transfer, as well as implementation of a finite-volume based numerical scheme. This will allow expanding the code's capabilities for realistic simulations of the chromospheric and coronal dynamics.

Overall, MANCHA3D is a versatile code that can be easily reconfigured for different scientific setups, from idealized to the realistic ones. Setting a simulation does not involve changing of the main code, but only requires from a user to modify a few external files specifying boundary and initial conditions, compiled together with the code, as well as the control file. Therefore MANCHA3D can be used by someone with relatively little experience in high-performance computing.

**Author Contribution** MM lead technical code development and optimization since 2018; wrote most of the text, build most of the figures and performed scaling tests reported in Section 4.

EK created the Mancha code, was its main developer till 2012, and supervised its development since; she wrote the Introduction, some parts of Sections 2.1, 2.2, 2.6 and 3.1, 3.3, 3.4, 3.5, 3.6, and produced Figure 5.

NV has created the modules of Mancha for realistic simulations of solar and stellar convection, contributed to overall code's development and parallelization; he wrote Sections 2.3-2.5.

AdV introduced the actual code's modular structure and parallelization, performed multiple optimizations, and contributed to writing Section 4.

AN programmed the heat conduction module and contributed to Section 2.1.

PAGM programmed the STS and HDS schemes and contributed to Sections 3.1.2-3.1.3.

MCV co-authored the creation of the Mancha code, contributed to its posterior development and simulation setups, and contributed to improving the paper text.

TF extended the code into 3D, participated in its parallelization, and contributed to improving the paper text.

DMG contributed with scientific simulations setups and improvements of the paper text.

PH contributed with scientific model for heat conduction module and contributed to improving the paper text.

ML improved 3D aspects of the code and contributed to improving the paper text.

MKP programmed the non-equidistant grid module and contribute to writing Section 3.2.2.

BPB contributed with code optimization, scientific simulations setups and contributed to improving the paper text.

APG contributed with scientific simulations setups, opacity binning techniques, and contributed to improving the paper text.

VL contributed with scientific simulations setups and contributed to improving the paper text.

IS contributed with scientific simulations setups and contributed to improving the paper text.

MMGM contributed with scientific simulations setups and contributed to improving the paper text.

**Funding** The authors thank the support from the European Research Council through the Consolidator Grant ERC-2017-CoG-771310-PI2FA and from the Agencia Estatal de Investigación del Ministerio de Ciencia e Innovación and the European Regional Development Fund (ERDF) through grant PID2021-127487NB-I00. The scaling tests were performed on the PizDaint supercomputer under the SOLARNET project. VL acknowledges support by the ERC Advanced Grant PROMINENT from the European Research Council under the European Union's Horizon 2020 research and innovation programme (grant agreement No. 833251 PROMINENT ERC-ADG 2018). BPB is supported be the FWO grant 1232122N. ML acknowledges support through the Ramón y Cajal fellowship RYC2018-026129-I from the Spanish Ministry of Science and Innovation, the Spanish National Research Agency (Agencia Estatal de Investigación), the European Social Fund through Operational Program FSE 2014 of Employment, Education and Training and the Universitat de les Illes Balears. This publication is part of the R+D+i project PID2020-112791GB-I00, financed by MCIN/AEI/10.13039/501100011033.

**Code Availability** MANCHA3Dcode is freely available via public Gitlab repository, https://gitlab.com/Mancha3D.

## Declarations

**Conflict of interest** The authors declare that they have no conflicts of interest.